\documentclass[a4paper,11pt]{article}
\pdfoutput=1 
\usepackage{jheppub}

\usepackage[utf8]{inputenc}

\usepackage{amssymb,amsmath,amsfonts}
\usepackage{mathtools}
\usepackage{mathrsfs}
\usepackage{bbm}
\usepackage{slashed}
\usepackage{nicefrac}
\usepackage{graphicx}
\usepackage{caption}
\usepackage{subcaption}
\usepackage{esint}

\usepackage{graphicx}
\usepackage[dvipsnames]{xcolor}
\usepackage{array}

\usepackage{simplewick}

\usepackage{hyperref}
\usepackage{xparse}
\usepackage{xspace}
\usepackage{soul}

\usepackage{tikz}
\usetikzlibrary{decorations.pathmorphing}
\usetikzlibrary{automata,positioning}

\usepackage{cancel}
\usepackage[normalem]{ulem}

\usepackage{xifthen}
\usepackage{dsfont}
\usepackage[titletoc]{appendix}
\usepackage{booktabs}
\usepackage{units}
\usepackage{braket}

\newcommand{\gettitle}{}
\hypersetup{linkcolor=black
	colorlinks,
	linkcolor={red!75!black},
	citecolor={blue!75!black},
	urlcolor={blue!75!black},
	pdftitle={\gettitle},
	pdfauthor={Barata, Mehtar-Tani},
	pdfkeywords={Perturbative QCD} {Jet quenching},
	bookmarksopen=true,
	bookmarksopenlevel=2,
	bookmarksnumbered=true
}
\makeatletter
\newcommand\makebig[2]{%
  \@xp\newcommand\@xp*\csname#1\endcsname{\bBigg@{#2}}%
  \@xp\newcommand\@xp*\csname#1l\endcsname{\@xp\mathopen\csname#1\endcsname}%
  \@xp\newcommand\@xp*\csname#1r\endcsname{\@xp\mathclose\csname#1\endcsname}%
}
\makeatother
\makebig{biggg} {1.3}

\def\bs{\boldsymbol} 
\def\del{\partial}
\def\bdel{\bs\partial}

\newcommand{\eqn}[1]{Eq.~\eqref{#1}}

\long\def\comment#1{ }

\newcommand{\nn}{\nonumber\\ }

\def\be{\begin{eqnarray*}}
\def\ee{\end{eqnarray*}}
\def\beq{\begin{eqnarray}}
\def\eeq{\end{eqnarray}}
\newcommand{\bea}{\beq \begin{aligned}}
\newcommand{\eea}{\end{aligned}\eeq}

\def\k{{\boldsymbol k}}

\def\q{{\boldsymbol q}}
\def\p{{\boldsymbol p}}

\def\x{{\boldsymbol x}}
\def\y{{\boldsymbol y}}
\def\z{{\boldsymbol z}}

\def\0{{\boldsymbol 0}}

\def\k{{\boldsymbol k}}

\def\x{{\boldsymbol x}}
\def\y{{\boldsymbol y}}
\def\p{{\boldsymbol p}}

\def\z{{\boldsymbol z}}

\def\khat{\hat P} 
\def\jhat{\hat K} 
\def\rhat{\hat R} 
\def\lambdaq{\lambda}
\def\vLO{v^\text{\tiny HO}}

\def\rme{{\rm e}}
\def\dd{\text{d}}

\def\and{ \quad\text{and}\quad}

\def\Re{\text{Re}}

\def\rmR{{\rm Re}}

\def\abar{{\rm \bar\alpha}}

\def\cP{{\cal P}}
\def\cK{{\cal K}}
\def\cM{{\cal M}}

\def\cO{{\cal O}}

\def\rmd{{\rm d}}

\def\dd{\text{d}}

\def\dd{\text{d}}

\def\abar{\bar\alpha}

\def\abar{\bar\alpha}

\begin{document}

\title{Medium-induced radiative kernel with the Improved Opacity Expansion}

\author[a]{Jo\~ao Barata,}
\emailAdd{joaolourenco.henriques@usc.es}
\affiliation[a]{Instituto Galego de Fisica de Altas Enerxias (IGFAE), Universidade de Santiago de Compostela,E-15782 Galicia, Spain}
\author[b,c]{Yacine Mehtar-Tani,}
\emailAdd{mehtartani@bnl.gov}
\affiliation[b]{Physics Department, Brookhaven National Laboratory, Upton, NY 11973, USA}
\affiliation[c]{RIKEN BNL Research Center, Brookhaven National Laboratory, Upton, NY 11973, USA}
\author[d]{Alba Soto-Ontoso,}
\affiliation[d]{Institut de Physique Théorique, Université Paris-Saclay, CNRS, CEA, F-91191, Gif-sur-Yvette, France}
\emailAdd{alba.soto@ipht.fr}
\author[e]{and Konrad Tywoniuk}
\emailAdd{konrad.tywoniuk@uib.no}
\affiliation[e]{Department of Physics and Technology, University of Bergen, 5007 Bergen, Norway}

\date{\today}
\abstract{
We calculate the fully differential medium-induced radiative spectrum at next-to-leading order (NLO) accuracy within the Improved Opacity Expansion (IOE) framework. This scheme allows us to gain analytical control of the radiative spectrum at low and high gluon frequencies simultaneously. The high frequency regime can be obtained in the standard opacity expansion framework in which the resulting power series diverges at the characteristic frequency $\omega_c\sim \hat q L^2$. In the IOE, all orders in opacity are resumed systematically below $\omega_c$ yielding an asymptotic series controlled by logarithmically suppressed remainders down to the thermal scale $T \ll \omega_c$, while matching the opacity expansion at high frequency. 
Furthermore, we demonstrate that the IOE at NLO accuracy reproduces the characteristic Coulomb tail of the single hard scattering contribution as well as the Gaussian distribution resulting from multiple soft momentum exchanges. Finally, we compare our analytic scheme with a recent numerical solution, that includes a full resummation of multiple scatterings, for LHC-inspired medium parameters. We find a very good agreement both at low and high frequencies showcasing the performance of the IOE which provides for the first time accurate analytic formulas for radiative energy loss in the relevant perturbative kinematic regimes for dense media.

}

\date{\today}
\maketitle
\flushbottom

\section{Introduction}\label{sec:intro}

High-energy collisions of heavy nuclei provide the necessary conditions for creating an extended medium of hot and dense nuclear matter, referred to as quark-gluon plasma (QGP). The appearance of a short-lived stage of this exotic state of matter leaves a strong imprint on particle production at all momentum scales and is quantified by high-precision experimental measurements. In this work, particular attention is devoted to the high-energy particles that traverse the medium and  can be used as perturbatively well controlled probes of the microscopic properties of the QGP~\cite{Gyulassy:1990ye,Wang:1991xy}. In terms of experimental observables, these objects emerge in the detectors as collimated sprays of particles and energy, colloquially referred to as \emph{jets}. Jet modifications, quantified with respect to a baseline obtained in proton-proton collisions (or ``vacuum''), have been intensively studied at RHIC~\cite{Adcox:2001jp,Adler:2002xw,Adler:2002tq} and LHC~\cite{CMS:2012aa,Aad:2012vca,Aad:2015wga,Aamodt:2010jd,Aad:2014bxa,Adam:2015ewa,Khachatryan:2016jfl} since more than a decade.

The suppression and modification of jets produced in heavy ion collisions, commonly known as \emph{jet quenching}, is driven by two main phenomena: transverse momentum broadening and energy loss. The former refers to the acquisition of transverse momentum by the highly energetic partons that make up the jet through elastic interactions with the medium following Brownian motion. This diffusion in momentum space is characterized by the transport coefficient: 
\beq
\hat q \equiv \frac{\rmd \langle k_\perp^2 \rangle_{\rm typ}}{\rmd t} \, ,
\eeq 
where $\langle k_\perp^2 \rangle_{\rm typ}$ is the typical squared transverse momentum transfer. An important role is also played by induced energy loss, such as caused by drag and inelastic, or radiative, processes. The latter component is a result of bremsstrahlung radiation triggered by collisions with medium constituents. The associated mean energy loss was found to scale as $L^2$, where $L$ is the length of the plasma, and thus constitutes an important, if not the dominant, source of jet quenching for large media, even though it is ``naively'' suppressed by a power of the coupling constant~\cite{BDMPS2}. 

The above physical picture applies for the regime of multiple scattering during the passage through the medium. This is the case when the opacity $\chi = L/\ell_{\rm mfp}$, defined as the ratio between the medium length, $L$, and the mean free path, $\ell_{\rm mfp}=(\rho\sigma_{\rm el})^{-1}$ (where $\rho$ is the density of scattering centers and $\sigma_{\rm el}$ the total elastic cross section), is of order one or larger, i.e. $\chi \gtrsim 1$. In effect, many interactions, i.e. those that occur within the formation time,  coherently participate in inducing gluon radiation, in an analogous way to the well-known Landau-Pomeranchuk-Migdal (LPM) effect in QED~\cite{LPM1,LPM2}. It was soon understood that neglecting these interference effects would fail in adequately describing the radiative spectrum in a substantial region of phase space in such media. Focusing on the diffusion approximation, and thereby neglecting the Coulomb tail that captures the physics of rare hard momentum transfers, the radiative spectrum could be analytically computed albeit with an ambiguity in setting the upper bound for the Coulomb logarithm in $\hat q$~\cite{BDMPS1,BDMPS2,BDMPS3,BDMPS4,BDIM1,Arnold:2020uzm}. 

More precise calculations can be performed in the dilute regime, where at most a few scatterings contribute and the full parton-medium interaction potential can be used~\cite{Gyulassy:2000fs,Wiedemann,Guo:2000nz,Vitev:2009rd,Chien:2015hda,Wiedemann,GLV}. However, their domain of applicability is limited to low opacity $\chi \ll 1$ or large momentum transfer. Due to the finite radius of convergence of the opacity expansion such an approach diverges~\cite{Arnold_simple} and thus cannot be applied in the case of a dense medium. 

As we have mentioned, in both low and high opacity regimes, under a set of physically motivated assumptions that will be revisited in what follows, the medium-induced spectrum can be computed analytically. However, the scenario explored in current experimental facilities, such as the LHC or RHIC, is expected to be the one where the jet undergoes a handful of scatterings, $\mathcal{O}(1-50)$, with the medium~\cite{Feal:2019xfl}. As such, neither of the above limiting forms is in its exact domain of applicability, thus hindering quantitative theory-to-data comparisons and the extraction of the medium parameters through the jet physics program in current colliders. 

These limitations in the analytic front have motivated the investigation of the radiative spectrum numerically~\cite{numerical1,Zakharov:2004vm,numerical2,numerical3,CarlotaFabioLiliana,Andres:2020kfg,Feal:2019xfl}. However, the proposed approaches are in general less transparent and potentially more computationally costly as compared to their analytic counterparts for computing jet observables where multiple gluon radiation is to be resummed to all orders~\cite{ASW2}, for applications to jet suppression~\cite{Mehtar-Tani:2021fud,Takacs:2021bpv}, or as a building block for Monte Carlo event generators~\cite{Lokhtin:2005px,Zapp:2008gi,Armesto:2009fj,Renk:2010zx,Young:2011ug,Casalderrey-Solana:2014bpa,He:2015pra,Cao:2017zih,Putschke:2019yrg,Caucal:2019uvr,Ke:2020clc}. 

A first step towards a more precise control over the accuracy of analytic calculations was first taken in Ref.~\cite{Arnold:2008zu}, where the next-to-leading logarithmic corrections to $\hat q $ were computed in the multiple soft scattering regime or, equivalently, in the infinite length medium limit. More recently, substantial progress was made in unifying the low and large opacity regimes while recovering the results of Ref.~\cite{Arnold:2008zu} in the soft regime. This new scheme, dubbed the Improved Opacity Expansion (IOE), has been shown to successfully meet this goal when applied to the description of the single particle broadening probability~\cite{broadening_paper} and to the medium-induced gluon energy spectrum~\cite{IOE1,IOE2,IOE3}, which constitute the major tools used in a multitude of well established phenomenological models of jet quenching~\cite{Hybrid,Saclay,Kutak1,Blanco:2020uzy,Soeren1,Soeren2,Adhya:2019qse}. In a nutshell, this framework is built as a series expansion of the in-medium scattering cross section where the zeroth order term encodes the multiple scattering solution and higher N$^n$LO orders\footnote{The nomenclature used here to denote the orders in the IOE should not be confused with the more familiar perturbative expansion in powers of the coupling constant.} 
in the series account for $n$ hard scatterings with the medium. 

This paper aims at computing the fully differential medium-induced spectrum at NLO accuracy in the IOE framework. For different values of the gluon energy and transverse momentum, ($\omega,\k$),\footnote{Bold letters denote 2D transverse vectors in this paper, while their modulus is written as $|\k|\equiv k_\perp$.}  we provide an analytic formula for an arbitrary medium profile that requires numerical integrations of the same order in computational complexity as the ones encountered in the multiple soft scattering limit~\cite{ASW1,ASW2}. In the simplified scenario in which the medium is treated as a brick of constant density, we have obtained closed, analytic formulas for the asymptotic behavior of the spectrum computed in the three different setups considered in this work: the IOE at NLO, the single-hard scattering approximation (${\rm SH}$) and the multiple soft scattering regime (${\rm MS}$). In addition, we make a phenomenologically oriented comparison with the all-orders numerical spectrum presented in Ref.~\cite{CarlotaFabioLiliana}. The numerical routines used in this publication are provided as ancillary files that can be found in Ref.~\cite{python_git}. 

The remainder of this paper is structured as follows. Section~\ref{sec:gluon_generic_IOE} revisits the Improved Opacity Expansion framework in full generality, including its application to the single particle momentum broadening and the medium-induced energy spectrum calculations. The core of this paper is Section~\ref{sec:IOE_radiation_spectrum}, where the fully differential spectrum is calculated with a high level of detail. Those readers more interested in the final result and not so much in the technicalities can find a summary of the ready-to-use formulas in Section~\ref{sec:summary}. Finally, numerical results for LHC-motivated medium parameters are presented in Section~\ref{sec:numerics}, including a comparison to BDMPS-Z, GLV and the resummed to all orders spectrum. Further details on the analytic calculations can be found in Appendices~\ref{app:cK_appendix}, \ref{app:Q}, \ref{app:vacuum-derivation} and \ref{app:In_Out_example}. 

\section{The Improved Opacity Expansion: an overview }
\label{sec:gluon_generic_IOE}

The Improved Opacity Expansion draws on the seminal 1948 work by Moli\`ere~\cite{Moliere}, where the transverse momentum broadening of charged particles in QED was described in such a way that the multiple soft scattering solution, well described by a Gaussian distribution, and the Coulomb power-law tail were reproduced in the appropriate limits. The IOE program consists in extending Moli\`ere's original approach to QCD and to more complex observables. So far, this strategy has been successfully applied to compute the medium-induced gluon energy spectrum~\cite{IOE1,IOE2,IOE3} and the transverse momentum broadening distribution of an energetic parton propagating through a dense QCD plasma~\cite{broadening_paper}.

As an introduction to the IOE, it will be instructive to first revisit how it applies to the two aforementioned observables: transverse momentum broadening and medium-induced gluon radiative spectrum. This will also serve us to lay the basis of the calculation of the fully differential spectrum.

\subsection{Transverse momentum broadening }
\label{sec:momentum-broadening}

The elementary in-medium process that underlies the observables that we discuss in this work is the elastic collision rate $\gamma_{\rm el}\equiv  \, \rmd \sigma_{\rm el}/\rmd^2\q$, where $\q\equiv (q_1,q_2)$ corresponds to the transverse momentum transfer in the $t$-channel between the hard probe and the medium. At leading order in the coupling the rate reads $\gamma_{\rm el} \sim  g^4n/q_\perp^{4}$, where $n$ corresponds to the density of scattering centers in the medium and the $1/q_\perp^4$ dependence denotes that, at short distances, the interaction is Coulomb-like. On the other hand, when $q_\perp \to 0$, the power law divergence should be screened by the medium at, roughly speaking, the Debye mass $m_D^2$ in the plasma.

Equipped with $\gamma_{\rm el}$, we can readily write a rate equation for the transverse momentum broadening distribution $\cP(\k;t)$, which gives the probability for a parton in color representation $R$ to acquire transverse momentum $\k$ due to in-medium propagation during a time $t$,
\beq
\label{eq:rate-eq}
 \frac{\del  \cP(\k,t)}{\del  t } = C_R\int_\q \,\gamma_{\rm el}(\q) \left[\cP(\k-\q,t) -\cP(\k,t) \right]\,,
\eeq
where the final time corresponds to the length of the medium $t=L$ and $C_R$ is the color factor associated to a representation $R$ of SU$(3)$.\footnote{In this paper we use the notation $\int_\q = \int \frac{\rmd^2 \q}{(2\pi)^2}$ to describe transverse momentum space integrals and $\int_\x= \int \rmd^2 \x$ for integration in position space.} The boundary condition at initial time $t=0$ is simply $\cP(\k,0) = (2\pi)^2 \delta^{(2)}(\k)$. The first term in Eq.~\eqref{eq:rate-eq} accounts for the gain in transverse momentum of the initial parton while the second term reflects the loss of probability for finding said parton with the measured momentum $\k$.  Notice that due to rotational symmetry the broadening probability is a function of the modulus of the transverse momentum vector, i.e. $k_\perp \equiv |\k|$. 

The integral of the collision rate yields the inverse mean-free-path between two collisions, i.e $\ell^{-1}_{\rm mfp} \equiv \int_\q\,  \gamma_{\rm el}(\q)$.  At low opacity, $\chi  \sim  L/\ell_{\rm mfp}  \ll 1$, the distribution is dominated by at most a single hard scattering (SH) and one finds 
\beq\label{eq:oe-lo-br}
 \cP^{\rm SH}(\k,L)= C_R\gamma_{\rm el}(\k)\,  L \sim   (4\pi)^2\frac{\alpha_s^2C_R nL}{k_\perp^4}\, .
\eeq
Conversely, at high opacity, multiple (soft) scatterings occur with order one probability and Eq.~\eqref{eq:rate-eq} can be approximated by a diffusion equation for which analytic solutions exist. This is done by expanding in gradients for $q_\perp\ll k_\perp$. The first non-vanishing contribution involves the jet quenching parameter $\hat{q}$, 
\beq 
\hat q = C_R\, \int^{q_{\rm max}}\frac{\rmd^2 \q }{(2\pi)^2} \, q_\perp^2 \, \frac{\rmd \sigma_{\rm el}}{\rmd^2 \q  } \approx 4\pi \alpha_s^2 C_R n(t)\log\frac{q_{\rm max}^2}{\mu_\ast^2} \,,
\eeq
where the integral over $\q$ is divergent in the ultraviolet and thus must be regulated, giving rise to the standard Coulomb logarithm, while the infrared region is cut-off by the screening mass that we denote by $\mu^2_\ast$.  Assuming $\hat q $ to be constant in time, the solution to the diffusion equation is a Gaussian~\cite{broadening_paper} and the associated broadening distribution  reads
\beq\label{eq:gaussian}
  \cP^{\rm MS}(\k,L) =\frac{4\pi }{\hat q  L } \rme^{-\frac{k_\perp^2}{
  \hat q L }} \, .
\eeq
Although this result describes the physics of multiple soft scattering (MS) of the probe in the medium, the diffusion approximation has two major drawbacks: (i) it misses the heavy $1/q_\perp^4$ tail associated with large momentum exchanges and (ii) the transport coefficient depends, logarithmically, on an undetermined ultraviolet cutoff scale.  

The IOE overcomes these two limitations by shifting the expansion point of the opacity scheme from the vacuum to the harmonic oscillator potential, resulting in the Gaussian distribution presented in Eq.~\eqref{eq:gaussian}. This shift in the expansion is easily performed in position space and thus we should consider the Fourier pair of $\cP(\k,t)$, 
\beq\label{eq:rate-momentum}
  \cP(\x,t) = \int_\k   \cP(\k,t) \, \rme^{i\x\cdot \k }\, .
\eeq
In position space, \eqn{eq:rate-eq} becomes local 
\beq\label{eq:rate-position}
 \frac{\del  \cP(\x,t)}{\del  t } = - \, v(\x) \cP(\x,t) \,,
\eeq
implying $ \cP(\x,t)=\rme^{-v(\x) t}$,
where the scattering potential $v(\x)$ combines the gain and loss terms and is thus ultraviolet finite 
\beq \label{eq:v-llog}
v(\x) = C_R\int_\q \, \gamma_{\rm el}(\q)  \left(1- \rme^{i\q\cdot \x}\right) \propto x_\perp^2 \log \frac{1}{x_\perp^2 \mu_\ast^2 } \, . 
\eeq
In this example, \eqn{eq:rate-position} can be directly integrated, but this is not generally possible as we shall see in the case of the radiative spectrum. Furthermore, one still needs to invert the Fourier transform and this is where the IOE scheme will be particularly useful as it allows us to reduce the Fourier transform to a sum of standard integrals. 

Let us recall the main difference between the Improved Opacity Expansion procedure and the usual Opacity Expansion (OE) strategy~\cite{Gyulassy:2000fs,Wiedemann,Guo:2000nz}. The latter performs an expansion directly in powers of $v(\x)$ of \eqn{eq:rate-position}, yielding a series in powers of $q_\perp^{-2}$ once introduced in \eqn{eq:rate-momentum}, with the leading contribution given by \eqn{eq:oe-lo-br}. In the IOE, one shifts the expansion point to be a solution to Eq.~\eqref{eq:rate-position} with the potential $v = \vLO\equiv \hat q  x_\perp^2/4$, whose Fourier transform can be carried out to yield \eqn{eq:gaussian}. If we denote such a solution by $\cP^{\rm LO}$, then the aforementioned shift of the expansion point leads to 
\begin{equation}
\cP(\x,L) =\big[ 1-  \delta v(\x) \big]\cP^{\rm LO}(\x,L) +\mathcal{O}(\delta v^2) \, . 
\end{equation}
 Here the scattering potential is split into two terms, i.e. $v= \vLO+\delta v$, such that
 $|\delta v| \ll |\vLO| $, in which case $\delta v$ can be regarded as a perturbation around the potential $\vLO$. In doing so we aim to tame the divergence of the plain Opacity Expansion series at low enough $\q$, typically when $q_\perp^2 < \hat q L$. This separation of $v$ into $\vLO$ and $\delta v$ is in general arbitrary and requires the introduction of a matching scale $Q$. Clearly, truncating  the IOE series at a fixed order introduces a residual dependence on the separation scale that is of the order of the remainder and thus can be safely neglected. It can nevertheless be used to gauge the uncertainty associated with the fixed order calculation very much like scale dependence encountered in \emph{standard} perturbation theory calculations. 
 
To illustrate this point consider the leading logarithmic form given in \eqn{eq:v-llog}. One would trivially write 
\begin{equation}\label{eq:ppp}
  x_\perp^2 \log \frac{1}{x_\perp^2 \mu_\ast^2 } = x_\perp^2 \left[ \log \frac{Q^2}{ \mu_\ast^2 }  + \log \frac{1}{ x_\perp^2 Q^2 }  \right] 
\end{equation}
  and define $\vLO \propto  x_\perp^2 \log \frac{Q^2}{ \mu_\ast^2 } $ and $ \delta v \propto x_\perp^2 \log \frac{1}{ x_\perp^2 Q^2 }  $, up to overall time-dependent factors. A natural candidate for the separation scale in the case of momentum broadening is $Q^2 \sim \hat q t$, corresponding to the average momentum squared accumulated by the probe due to multiple soft momentum exchanges with the medium. In general, when considering other observables, the LO provides guidance to what scale should be chosen for $Q^2$.  We shall see below how to make this observation more precise, in particular regarding the ultraviolet behavior of $\hat q$. 

Not only the IOE allows to fix the diverging behavior of the Opacity Expansion, but also provides a good approximation of the exact result at low transverse momentum provided the following hierarchy of scales is met
\beq
Q^2 \gg \mu_\ast^2 \,.
\eeq
This ensures that the Coulomb logarithm is large, i.e. $\log (Q^2/\mu_\ast^2) \gg 1$, and since at low $k_\perp$, the $\x$ integral is dominated by the region $x_\perp^2 \sim 1/Q^2$, we also have that  $\log \frac{1}{ x_\perp^2 Q^2 } \sim 1$. On the other hand, at large $k_\perp$ the rapidly oscillating Fourier phase implies that $x_\perp \ll  k_\perp^{-1} \ll  Q^{-1}$ which flips the relative order of the LO and its correction, i.e. $| \delta v| \gg |\vLO| $, and thus the logarithmic function in $v(\x)$ can no longer be neglected. Since large momentum transfers are associated with steeply falling cross sections, such a case is associated with rare hard scatterings in the medium, and perturbation theory is applicable, recovering the standard Opacity Expansion.

Following this more qualitative discussion that highlights the strengths of the IOE approach, let us make the discussion more quantitative and rigorous by recalling some of the results presented in Ref.~\cite{broadening_paper}. First, in jet quenching phenomenology two models for the in-medium scattering rate are typically considered. One option, referred to as Gyulassy-Wang (GW) model~\cite{GW}, is to describe the medium as an ensemble of static scattering centers with Yukawa like potentials, with the in-medium rate given by
\beq\label{eq:GW}
\gamma_{\rm el}^{\rm  GW} (\q,t)= \frac{g^4 n(t)}{(q_\perp^2+\mu^2)^2} \, ,
\eeq
where $\mu$ is the GW screening mass and $n$ the density of scattering centers in the medium. This leads, see Eq.~\eqref{eq:v-llog}, to a scattering potential of the form
\begin{align}
\label{eq:v_GW_text}
    v^{\rm GW}(\x,t)&=\frac{\hat{q}_0(t)}{\mu^2} \big[ 1-\mu x_\perp K_1(\mu x_\perp)\big] \, ,
\end{align}
where we have introduced the \textit{bare} jet quenching parameter $\hat{q}_0(t)=4\pi \alpha_s^2 C_A n(t)$ and $K_1$ is the modified Bessel function of the second kind of order $1$. Another popular choice is to describe the medium as a thermal bath, so that the scattering potential can be perturbatively computed using Hard Thermal Loop (HTL) effective theory~\cite{HTL}, with
\beq\label{eq:HTL}
\gamma_{\rm el}^{\rm  HTL} (\q,t)= \frac{g^2m_D^2(t)T}{q_\perp^2\,\big[q_\perp^2+m_D^2(t) \big]} \,.
\eeq
Here $T$ is temperature of the medium and $m_D$ the Debye screening mass. The corresponding scattering potential reads
\begin{align}
\label{eq:v_HTL_text}
v^{\rm HTL}(\x,t)= \frac{2\hat{q}_0(t)}{m_D^2(t)}\left[ K_0(m_D(t)x_\perp)+\log\left(\frac{m_D(t)x_\perp}{2}\right)+\gamma_E \right] \, ,
\end{align} 
where now $\hat{q}_0(t)=\alpha_s C_A m_D^2(t) T$, $\gamma_E = 0.577216\ldots$ is the Euler-Mascheroni constant and $K_0$ is the modified Bessel function of the second kind of order $0$.\footnote{Here we defined the jet quenching parameter for gluons, i.e. $C_R=C_A$.} The differences and similarities between these two models have been extensively discussed in Refs.~\cite{IOE3,broadening_paper}. To leading logarithmic order, they can be unified in an universal form, in accordance with \eqn{eq:v-llog}, 
\begin{equation}\label{eq:v_LL}
v(\x,t)\equiv  \frac{1}{4} \hat{q}_0(t) x_\perp^2 \log \frac{1}{x_\perp^2\mu_\ast^2} +\cO(x_\perp^4\mu_\ast^2)\, ,
\end{equation}
where $\mu_\ast$ is a universal screening mass that can be mapped to the masses of both models considered above.\footnote{The GW mass $\mu$ is related to the universal mass $\mu_\ast$ by $4\mu_\ast^2=\mu^2 \rme^{-1+2\gamma_E}$, and the Debye mass $m_D$ in HTL corresponds to $4\mu_\ast^2= m_D^2 \rme^{-2+2\gamma_E} $ \cite{broadening_paper}.}

Applying the IOE prescription to split the potential as $v=\vLO+\delta v$, see \eqn{eq:ppp}, and inserting it back into \eqn{eq:rate-position}, we obtain, after expanding in powers of the perturbative potential $\delta v$,
\begin{equation}\label{eq:expli_cP_series}
\begin{split}
\cP(\k,L)&=\int_\x \, \rme^{-i \x \cdot \k }\rme^{-\frac{1}{4}x_\perp^2Q^2}\sum_{n=0}^{n_{\rm max}}  \, \frac{(-1)^n Q_{s0}^{2n}}{4^nn!} \, x_\perp^{2n} \log^{n}\frac{1}{x_\perp^2Q^2} \\
&\equiv \cP^{\rm LO}(\k,L) + \cP^{\rm NLO}(\k,L) + \cP^{\rm  NNLO}(\k,L) + \ldots \, ,
\end{split}
\end{equation}
where we identify the next-to-leading order (NLO) term with the contribution $\mathcal{O}(\delta v)$, the next-to-next-to-leading order (NNLO) with the $\mathcal{O}(\delta v^2)$ term, and so on.\footnote{The series is truncated at $n_{\rm max}\sim Q_{s0}^2/\mu_\ast^2$ since formally this is a divergent asymptotic series; the divergence is physically associated to the fact that $x_\perp$ can not be smaller than $1/\mu_\ast$ --- see Ref.~\cite{Iancu:2004bx} for a further discussion on this truncation.} In Eq.~\eqref{eq:expli_cP_series}, we have introduced the \emph{bare} saturation scale 
\begin{align}\label{eq:old_Qs0}
  &Q_{s0}^2(L) = \int_{0}^{L} \rmd t\, \hat q_0(t) \, ,
 \end{align}
where we allow the \textit{bare} jet quenching parameter to vary in time. In addition, we define the \textit{effective} jet quenching parameter $\hat{q}(t)=\hat{q}_0(t) \log\frac{Q^2}{\mu_\ast^2}$, where the logarithmic dependence appears naturally from the splitting of $v(\x)$. As discussed above, the definition of the matching scale, $Q$, can not be cast in a closed form, since it enters the definition of $\hat{q}$ as well as depends on it directly. In turn, it is obtained by solving the transcendental equation
\begin{align}\label{eq:old_Qb}
  &Q_b^2\equiv Q_s^2(L) = \int_0^{  L} \rmd t\, \hat q_0(t)   \,\log \frac{Q_b^2( L)}{\mu_\ast^2}\, ,
 \end{align}
where, following our previous reasoning, we have identified $Q^2 \equiv Q^2_b$ with the \emph{effective} saturation scale $Q_s^2$.
We truncate \eqn{eq:expli_cP_series} at NLO accuracy, since already at this order both the hard and soft regimes should be well described. The resulting broadening distribution reads~\cite{broadening_paper}
\begin{equation}\label{eq:golden}
\cP^{\rm{LO+NLO}}(\k,L)=  \frac{4\pi}{Q_s^2} \rme^{-x} - \frac{4\pi}{Q_s^2} \lambdaq \left\{1-2 \rme^{-x} + \left(1-x\right) \left[{\rm Ei}\left( 4x\right)-\log 4x\right] \right\}\, ,
\end{equation}
where $x = k_\perp^2/Q_s^2$, and
\beq
\lambdaq \equiv \frac{\hat{q}_0}{\hat{q}}=\frac{1}{\log \frac{Q^2}{\mu_\ast^{ 2}}} \ll 1\,,
\eeq
is the expansion parameter of the series in the regime $k_\perp^2 \lesssim Q_s^2$.\footnote{The exponential integral function is defined as ${\rm {Ei}}(x)=\displaystyle\int_{-\infty}^x \rmd t \, \frac{\rme^{t}}{t}$.} At large momentum exchanges, $k_\perp^2 \gg Q_s^2$, one obtains from the NLO term, and in accordance with \eqn{eq:oe-lo-br}, that 
\begin{align}\label{eq:plot_ppp}
\cP(\k,L)^{\rm{NLO}}\Big\vert_{k_\perp^2\gg Q_{s}^2}=4\pi\frac{Q_{s0}^2}{k_\perp^4}+\mathcal{O}\left(\frac{Q_{s0}^4}{k_\perp^6}\right) \, ,
\end{align}
while the LO term is exponentially suppressed. In this high momentum limit, we recover the Coulomb tail encoded in a single scattering in the medium. On the other end, when $k_\perp^2 \ll Q_s^2$, we find
\beq
\label{eq:assist_1}
\cP(\k,L)^{\rm{LO+NLO}}\Big\vert_{k_\perp^2\ll Q_{s}^2} =  \frac{4\pi}{Q_s^2}\left( 1+ \lambdaq\log 4 \rme^{1-\gamma_E} \right)+ \cO\left( \lambdaq^2\right)\, .
\eeq
The first term corresponds to the LO contribution. Thus, the NLO term, up to a small constant logarithm, is of the same functional form as the LO but power suppressed by $\lambdaq \ll 1$. In fact, one can show that, in this regime, perturbative corrections in the IOE scale as the LO term, each increasing order suppressed by an extra power of $\lambdaq=\frac{\hat{q}_0}{\hat{q}}$. Hence, in this limit the LO term dominates and one recovers the multiple soft solution, which correctly describes the physics at play.

\begin{figure}[t!]
  \centering
  \includegraphics[scale=.8]{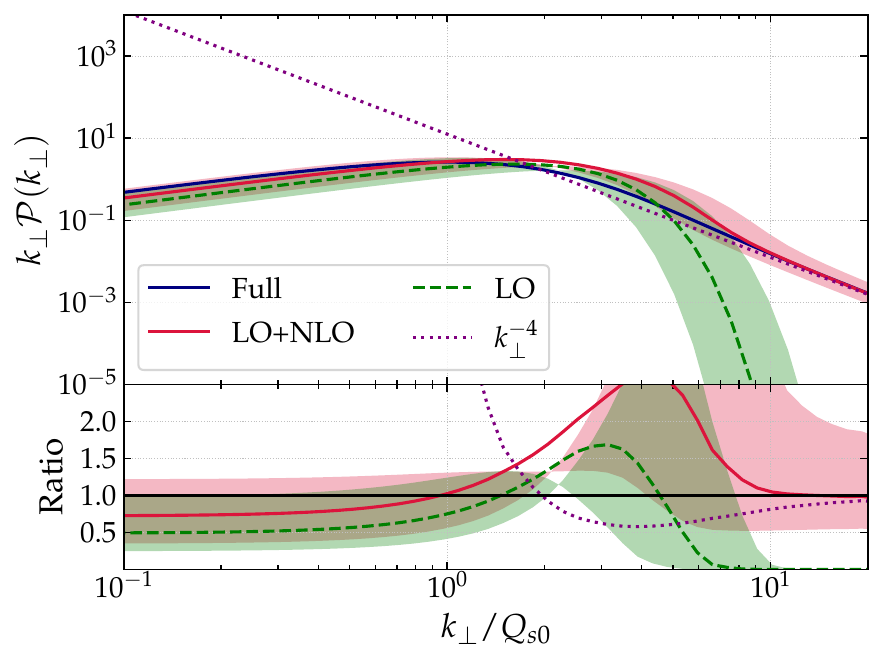}
  \caption{Comparison between the broadening probability distribution for the IOE at LO (dashed, green), at LO+NLO (solid, red) and the exact GW model result (solid, navy). In addition, we provide the single hard scattering solution given by \eqn{eq:plot_ppp}, which we denote by $k^{-4}_\perp$(dotted, purple). The ratio to the full solution is presented in the bottom panels. The uncertainty band arises from variations in the matching scale by factors of $2$ and $1/2$. The medium parameters are $\hat q_0\!=\!0.16$~GeV$^3$, $L=6$~fm and $\mu_\ast=0.355$~GeV. They are identical to the ones used in Section~\ref{sec:numerics}.}
  \label{fig:broad}
\end{figure}

In Fig.~\ref{fig:broad} we numerically compare the broadening distribution $\cP(\k,L)$, for a medium with constant $\hat{q}_0$, computed up to LO and NLO in the IOE, with the full $\cP$ obtained using Eqs.~\eqref{eq:rate-momentum}, \eqref{eq:rate-position} and the GW potential in \eqn{eq:v_GW_text}. The result follows the above discussion: at large momentum transfers, $k_\perp^2\gg \hat{q}L$, the NLO term dominates and converges to the full result dominated by the single hard scattering result ($k^{-4}_\perp$). On the other hand, at low momentum transfers the LO and LO+NLO become comparable, reproducing the full result within an uncertainty band associated to the remaining freedom in the definition of $Q_b^2$. The biggest mismatch between the LO+NLO result and the full distribution happens near the peak of the distribution and could be eventually improved by adding more orders in the series. Nonetheless, it is clear that the IOE approach provides a neat interpolation between the soft and hard regime, instead of properly describing just one of these regions.

\subsection{The energy spectrum}
\label{sec:energy-spectrum}

As a second illustrative example, we consider the application of the IOE to compute the medium-induced gluon energy spectrum. The in-medium emission spectrum of a soft gluon with energy $\omega$ from a hard parton with energy $E\gg \omega$ in color representation $R$ can be compactly cast as \cite{Blaizot:2015lma} 
\begin{align}
\label{eq:BDMPS_spec_energy}
\omega\frac{\rmd I}{\rmd \omega } =\frac{2\bar{\alpha} \pi}{\omega^2} \Re \int_0^\infty \rmd t_2 \int_0^{t_2} \rmd t_1  \,
\bdel_\y\cdot \bdel_\x \big[\cK(\x,t_2;\y,t_1) - \cK_0(\x,t_2;\y,t_1)\big]_{\x=\y=0} \, .
\end{align}
Here $\bar{\alpha}=\alpha_sC_R/\pi$ and $\cK(\x,t_2;\y,t_1)$ is an effective emission kernel describing the broadening of the emitted gluon during its formation. It corresponds to the evolution operator of a quantum particle immersed in the imaginary potential $iv(\x)$ in 2+1 dimensions and obeys the Schr\"odinger equation 
\begin{equation}
\label{eq:cK_Sch}
\left[i\frac{\partial}{\partial t}+\frac{\bdel^2_\x}{2\omega}+iv(\x,t)\right]\cK(\x,t;\y,t_1)=i\delta^{(2)}(\x-\y)\delta(t-t_1) \,,
\end{equation}
which resums multiple scatterings of the radiated gluon with the medium between the emission times $t_1$ and $t_2$ in the amplitude and its complex conjugate, respectively.

For a general potential $v(\x,t)$ that includes the Coulomb tail at large momentum transfers, a closed form solution to Eq.~\eqref{eq:cK_Sch} is not known. An analytical solution can nevertheless be obtained for two special choices of the potential: vacuum and harmonic oscillator. In the vacuum case, setting $v(\x,t)=0$ leads to the following solution of Eq.~\eqref{eq:cK_Sch}
\begin{align}\label{eq:cK_vac}
\cK_0(\Delta\x,\Delta t) = \frac{\omega }{2\pi i \Delta t} \exp\left(i\frac{\omega\Delta\x^2}{2 \Delta t}\right) \, ,
\end{align}
where $\Delta\x = \x-\y$ and $\Delta t = t_2-t_1$. Note that this contribution is explicitly removed in Eq.~\eqref{eq:BDMPS_spec_energy} so that the result is only sensitive to the purely medium-induced contribution. In fact, we can also express the resummed propagator, given by the solution of Eq.~\eqref{eq:cK_Sch}, as a Dyson-like iterative equation that resums multiple interactions around the vacuum solution, namely
\begin{align}
\label{eq:cK-opacity-expansion}
\cK(\x,t_2;\y,t_1) &= \cK_0(\x-\y, t_2-t_1) 
- \int_{t_1}^{t_2} \rmd s \int_{\z} \, \cK_0(\x-\z,t_2 -s) v(\z,s) \cK(\z,s; \y,t_1) \,.
\end{align}
From the structure of the equation, we immediately see that, for a time independent rate $v(\x,t) = v(\x)$, the function $\cK$ only depends on $\tau \equiv t_2-t_1$. This equation is equivalent to an expansion in medium opacity $\chi$, defined as $\chi = L/\ell_{\rm mfp}$. Computing the radiative spectrum by truncating the expansion in \eqn{eq:cK-opacity-expansion} at a fixed order in $v(\x,t)$, or $\chi$, corresponds to the Opacity Expansion introduced in the previous section. Consistently, the $n-$th term in the OE scales as $\rmd I^{n}/\rmd \omega \sim \mathcal{O} \big( \chi^n \big)$. The single scattering solution corresponds to the $n=1$ truncation of the expansion and is often referred to as the GLV spectrum~\cite{GLV,Wiedemann}.

The other special case where Eq.~\eqref{eq:cK_Sch} is analytically solvable is when $v(\x,t) = \vLO(\x,t) = \frac{1}{4}\hat q(t) x_\perp^2$, that is, when the potential reduces to that of an harmonic oscillator. We recall that the IOE splits the leading logarithmic potential given in \eqn{eq:v_LL} as
\begin{equation}\label{eq:v_IOE}
  v(\x,t)\equiv \vLO+\delta v =\frac{1}{4} \hat{q}_0(t) x_\perp^2 \log \frac{Q^2}{\mu_\ast^2}+\frac{1}{4} \hat{q}_0(t)x_\perp^2 \log \frac{1}{x_\perp^2Q^2} \, ,
\end{equation}
where $Q$ is for now an undetermined matching scale, different from the one used for the broadening case. Yet, the \textit{effective} jet quenching parameter is $\hat q(t) = \hat q_0(t) \log Q^2/\mu_\ast^2$. Thus, similarly to transverse momentum broadening discussed in the previous section, the solution to Eq.~\eqref{eq:cK_Sch} with a quadratic potential, that we denote as $ \cK = \cK^{\rm LO}$, corresponds to the leading order (LO) term in the Improved Opacity Expansion. It reads
\begin{align}
\label{eq:cK_BDMPS_2}
 \cK^{\rm LO}(\x,t_2;\y,t_1)&= \frac{\omega}{2\pi i S(t_2,t_1)}\exp\left( \frac{i\omega}{2S(t_2,t_1)} \left[ C(t_1,t_2)\,\x^2+C(t_2,t_1)\,\y^2-2 \x\cdot\y\right] \right) \,.
\end{align}
Here, $C(t_2,t_1)$ and $S(t_2,t_1)$ are purely time dependent functions which are solutions to the initial condition problems~\cite{Arnold_simple}
\begin{equation}
\label{eq:cs-ho-equations}
\begin{split}
&\left[\frac{\rmd^2}{\rmd^2 t}+\Omega^2(t)\right]S(t,t_0)=0 \, ,\quad S(t_0,t_0)=0 \,,\quad \partial_t  S(t,t_0)_{t=t_0}=1 \, , \\   
&\left[\frac{\rmd^2}{\rmd^2t}+\Omega^2(t)\right]C(t,t_0)=0 \, ,\quad C(t_0,t_0)=1 \,,\quad \partial_t C(t,t_0)_{t=t_0}=0\, ,
\end{split}
\end{equation}
with the complex harmonic oscillator frequency $\Omega(t)$ given by 
\begin{equation}
\Omega(t)=\frac{1-i}{2}\sqrt{\frac{\hat{q}(t)}{\omega}} \, .
\end{equation}
More details on the properties of these functions can be found in Appendix~\ref{app:cK_appendix}.

Inserting \eqn{eq:cK_BDMPS_2} back into Eq.~\eqref{eq:BDMPS_spec_energy} and performing the time integrals, one obtains the spectrum at leading order in the IOE (or equivalently in the harmonic approximation). The final expression reads,
\begin{equation}
\label{eq:dIdw_BDMPS}
\omega\frac{\rmd I^{\rm LO}}{\rmd \omega} = 2\Bar{\alpha}\log \big\vert C(0,L) \big\vert\,,
\end{equation}
and is often referred to as the BDMPS-Z spectrum \cite{BDMPS3,BDMPS2}. The LO contribution to the IOE spectrum takes a particularly simple form in the case where the medium has an extension $L$ with a constant density $n$; we refer to this simple medium model as the plasma brick model. In the brick model one can simply define the jet quenching parameter as $\hat q(t) =\hat q\, \Theta(L-t) $, which allows one to write the $C$ and $S$ functions as
\beq 
\label{eq:S-and-C-functions}
S(t_2,t_1) = \frac{1}{\Omega} \sin \Omega (t_2-t_1) \,, \qquad \text{and} \qquad C(t_2,t_1)= \cos \Omega (t_2-t_1) \,.
\eeq
In this case, the well-known behavior of the spectrum at asymptotically at low and high frequencies is 
\beq
\label{eq:dIdw_BDMPS_cases}
\omega \frac{\rmd I^{\rm LO}}{\rmd \omega} \simeq   2\bar \alpha
 \begin{dcases}
\,\,\sqrt{\frac{\omega_c}{2\omega}}   \quad\qquad\text{for}\quad \omega \ll \omega_c\\
\,\,\frac{1}{12} \left(\frac{\omega_c}{\omega} \right)^2\quad\text{for} \quad \omega \gg \omega_c \,,\\
\end{dcases}
\eeq
where the characteristic gluon energy $\omega_c = \hat q L^2/2$ corresponds to gluons with maximal formation time, i.e. $
t_{\rm f} = L$. The behaviour in the soft limit highlights the Landau-Pomeranchuk-Migdal (LPM) interference~\cite{LPM1,LPM2} that occurs since the gluon is formed over timescales involving multiple interactions with the medium. The strong suppression at high gluon energies follows directly from the approximation of multiple soft interactions, implicit in the harmonic form. At these frequencies, i.e. $\omega > \omega_c$, the contribution from a single, hard scattering can be shown to dominate, as we will discuss below.

Let us now construct the contributions to the IOE beyond the LO term.  Adopting the decomposition provided by \eqn{eq:v_IOE} that allows us to separate the harmonic part from the $\x$ dependent Coulomb logarithm, and in analogy to the resummation around the vacuum solution given by Eq.~\eqref{eq:cK-opacity-expansion}, the full kernel can be written as
 \begin{equation}
 \label{eq:cK_ful_IOE}
\cK(\x,t_2;\y,t_1) = \cK^{\rm LO}(\x,t_2;\y,t_1)- \int_{t_1}^{t_2} ds \int_\z \, \cK^{\rm LO}(\x,t_2;\z,s)\, \delta v(\z,s)\cK(\z,s;\y,t_1) \,.
\end{equation}
Truncating this relation at $\mathcal{O}(\delta v^2)$, it is easily seen that the LO kernel is given by $\cK^{\rm LO}$ in Eq.~\eqref{eq:cK_BDMPS_2}. The NLO kernel reads
\begin{align}
\cK^{\rm NLO}(\x,t_2;\y,t_1) &= - \int_{t_1}^{t_2} \rmd s \int_\z \, \cK^{\rm LO}(\x,t_2;\z,s) \delta v(\z,s) \cK^{\rm LO}(\z,s;\y,t_1) \, ,
\end{align}
which can be used in Eq.~\eqref{eq:BDMPS_spec_energy} to compute the NLO contribution to the IOE spectrum, as was done for the LO term. Like in the broadening case, we do not consider higher order terms since truncating the series at NLO is enough to reproduce the single hard and multiple soft regimes. At this order, the spectrum reads~\cite{IOE1,IOE2,IOE3}
\beq
\label{eq:dIdw_LO_p_NLO}
\omega\frac{\rmd I^{{\rm LO+NLO}}}{\rmd \omega} = 2\bar{\alpha}\log \big\vert C(0,L) \big\vert +\frac{1}{2}\bar{\alpha}\hat{q}_0 \,\Re \int_0^L \rmd s\, \frac{-1}{k^2(s)} \log\frac{-k^2(s)}{Q^2 \rme^{-\gamma_E}} \,,
\eeq
where 
\begin{equation}\label{eq:k_NLO}
k^2(s)= -\frac{i\omega}{2} \left[{\rm Cot}(s,\infty)+{\rm Cot}(0,s)\right] \, .
\end{equation}
and we defined the ratio ${\rm Cot}(t_2,t_1) \equiv C(t_1,t_2)/S(t_2,t_1)$.

We now analyze the asymptotic forms of the spectrum by considering the brick model. In this case \eqn{eq:k_NLO} reduces to
\begin{equation}
\label{eq:k_NLO_brick}
k^2(s) = \frac{i\omega \Omega}{2} \big[ {\cot}\,\Omega s - \tan\Omega(L-s) \big] \, .
\end{equation}
At high frequencies, that is $\omega \gg \omega_c$ or $\Omega L\ll1$, one finds that Eq.~\eqref{eq:k_NLO_brick} leads to $k^2(s)\simeq i\omega/(2s)$. The high-frequency behavior of the NLO term is given by
\begin{equation}
\label{eq:NLO_scaling_high_energy}
\omega \frac{\rmd I^{\rm NLO}}{\rmd \omega}\simeq \Bar{\alpha}\hat{q}_0\frac{\pi}{4}\frac{L^2}{2\omega}
=\frac{ \Bar{\alpha}  \pi}{4} \chi\,\frac{\bar{\omega}_c}{\omega} \, .
\end{equation}
It dominates the spectrum, given the quadratic $\omega$ suppression of the LO term, see Eq.~\eqref{eq:dIdw_BDMPS}. In this last equation, we recall that the medium opacity parameter is $\chi \equiv \hat{q}_0 L / \mu_\ast^2 \sim L/\ell_{\rm mfp}$ and we introduced the high energy cut frequency  $\Bar{\omega}_c=\frac{1}{2}\mu_\ast^2 L$. Higher-order terms are all suppressed by at least one additional power of $1/\omega$ as well~\cite{IOE3}. Thus, similar to the discussion for the broadening distribution $\cP(\k)$, one observes that the dominant term, given in Eq.~\eqref{eq:NLO_scaling_high_energy}, comes solely from the NLO contribution and it can be shown to exactly match the medium-induced spectrum obtained by considering a single hard scattering in the medium~\cite{IOE1,GLV}, i.e. $n=1$ in the traditional Opacity Expansion. Furthermore, Eq.~\eqref{eq:NLO_scaling_high_energy} is independent of the matching scale, analogous to what was observed for $\cP(\k)$ in \eqn{eq:plot_ppp}.

At low frequencies, i.e. for $\omega \ll \omega_c$ or $\Omega L \gg 1$, the NLO term, containing the single hard scattering physics, can be simplified by noticing that $k^2(s) \simeq -\omega \Omega$, leading to~\cite{IOE1,IOE2,IOE3}
\begin{equation}
\label{eq:NLO_smallw_maineq}
 \omega \frac{\rmd I^{\rm NLO}}{\rmd\omega} 
\simeq \Bar{\alpha} \lambdaq \sqrt{\frac{\omega_c}{2\omega}}\left[\gamma_E+\log\left(\frac{\sqrt{\omega \hat{q}}}{\sqrt{2}Q^2}\right)+\frac{\pi}{4}\right] \,,
\end{equation}
which is equivalent to the next-to-leading logarithmic result derived in Ref.~\cite{Arnold:2008zu}. Again, as observed for the broadening distribution, in the soft regime higher order terms in the IOE scale as the LO contribution, see \eqn{eq:dIdw_BDMPS_cases}, with increasing power suppression by 
$\lambdaq = \big(\log Q^2/\mu_\ast^2 \big)^{-1} \ll 1$. 
In fact, one can show that
\begin{equation}
\label{eq:expansion_IOE_small_frequency}
\left.\frac{\rmd I/\rmd \omega}{\rmd I^{\rm LO}/\rmd \omega} \right|_{\omega \ll \omega_c} = 1+\lambdaq \left(a_0+a_1\log\frac{\sqrt{\omega \hat{q}}}{Q^2} \right)
   + \lambdaq^2 \left(b_0 + b_1 \log\frac{\sqrt{\omega \hat{q}}}{Q^2} + b_2\log^2\frac{\sqrt{\omega \hat{q}}}{Q^2}\right) +\ldots \, , 
\end{equation}
where $\{a_i, b_i\}$ are purely numerical coefficients \cite{IOE3}. This result implies that, in the soft limit, the full spectrum can be written in terms of the LO result with an effective jet quenching coefficient $\hat{q}_{\rm eff}$ that absorbs the additional logarithmic dependencies. 

More importantly, Eq.~\eqref{eq:expansion_IOE_small_frequency} imposes further constraints on the matching scale $Q^2$. To see this, let us first assume that the matching scale associated to the radiation spectrum 
$Q^2 = Q_r^2$ is independent of $\omega$. Then, in \eqn{eq:expansion_IOE_small_frequency} the logarithms in the denominators would be frozen. However, the numerator logarithms would evolve quite rapidly for $\mu_\ast^2\ll \omega^2 \ll (\hat{q}^2Q_r^4)^{-1}$, leading to a divergent series  (notice that the LO contribution would be negligible in this case). Thus, one concludes that $Q_r =  Q_r(\omega)$ in order for the spectrum to be free of unphysical divergences. In addition, one sees that the natural way to regulate the numerators is to take\footnote{Here we assume that $\hat{q}$ is time independent to simplify the discussion. The generic form for \eqn{eq:Q_r} is discussed in the next section.}
\begin{equation}\label{eq:Q_r}
  Q_r^2=\sqrt{\hat{q}_0\omega\log{\frac{Q_r^2}{\mu_\ast^2}}} \, .
  \end{equation}
Moreover, it can be shown~\cite{IOE3} that this form follows directly from the fact that once all orders in the IOE are resummed the spectrum takes the functional form of the LO term. For the present paper and the following calculations, the main message is that Eq.~\eqref{eq:Q_r} ensures that at low energies the spectrum is well behaved and non-physical divergences are absent. Also, and again in analogy to the broadening, at leading logarithmic order $Q_r^2 \sim \sqrt{\hat{q}_0\omega}$, which using the above relations for the gluon formation time and the average accumulated momentum, can be translated into the typical momentum acquired by a gluon with frequency $\omega \ll \omega_c$. The solutions of Eq.~\eqref{eq:Q_r} are discussed in Appendix~\ref{app:Q}.

Finally, we still need to ensure that $Q_r^2\gg\mu_\ast^2$ in order to justify the expansion. Ignoring the logarithmic dependence in the matching scale, we observe that the IOE approach only works if
\begin{equation}
\omega_{\rm BH} \ll \omega \,,    
\end{equation}
where we defined the characteristic Bethe-Heitler (BH) frequency as $\omega_{\rm BH} = \mu_\ast^4/\hat{q}_0$.
This condition means that the current scheme is not valid in the BH regime~\cite{Bethe:1953va}, see Ref.~\cite{Andres:2020kfg} for a similar conclusion and further discussion regarding the analytic treatment of the BH region. This regime is characterized by gluons with a formation time of the order of the mean free path in the medium, acquiring a momentum $k_\perp^2\sim \hat{q}_0 \ell_{\rm fmp} \sim \mu_\ast^2$ and with a typical energy $\omega_{\rm BH} \sim T$ of the order of the medium temperature. At this scale, non-linear dissipation effects take place \cite{Baier:2000sb} such as gluon absorption. However, in the case of large or dense enough media (such that $Q^2 \gg m_D^2$) the BH regime is power suppressed and radiative energy loss is dominated by frequencies in the deep LPM regime in the calculation of inclusive jet observables \cite{Baier:2001yt}.

\begin{figure}[t!]
  \centering
  \includegraphics[scale=0.8]{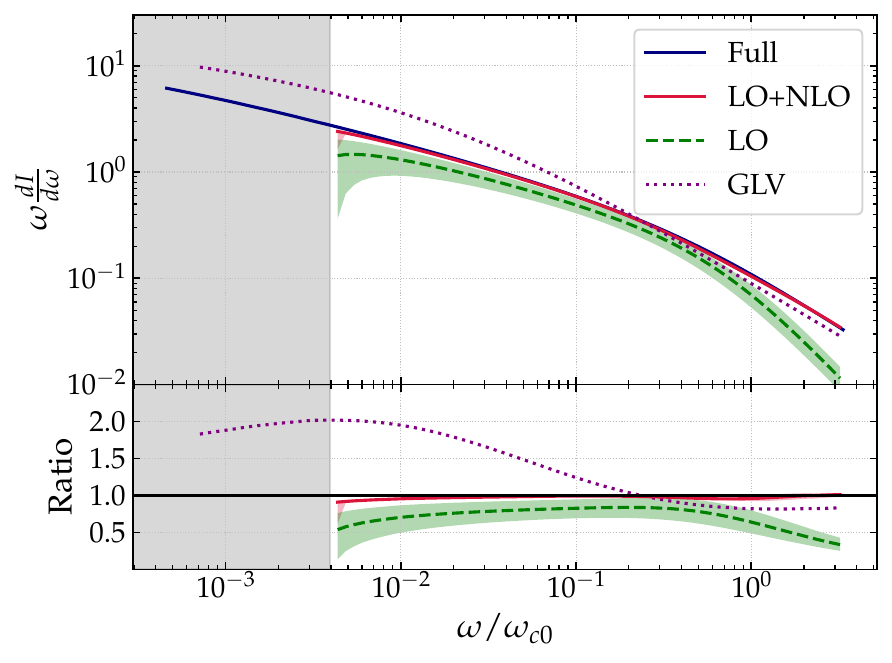}
  \caption{Comparison between the energy spectrum computed with GLV (dotted, purple), the IOE at LO (dashed, green), at LO+NLO (solid, red) and the all-order spectrum (solid, navy) as computed in~\cite{CarlotaFabioLiliana}. The ratio to the full solution is presented in the bottom panels. The uncertainty band arises from variations in the matching scale and the gray region indicates the regime in which Eq.~\eqref{eq:Q_r} does not have a solution. The parameters used are identical to those of Fig.~\ref{fig:broad} and $\omega_{c0}\!\equiv\!\hat q_0 L^2$.}
  \label{fig:spectrum}
\end{figure}
In Fig.~\ref{fig:spectrum}, we compute the medium-induced single gluon spectrum up to NLO in the IOE, comparing with a full numerical solution to Eq.~\eqref{eq:BDMPS_spec_energy}~\cite{CarlotaFabioLiliana} and the GLV spectrum, corresponding to the limit of single scattering in the medium. The gray band indicates the region in which  Eq.~\eqref{eq:Q_r} does not have a valid solution, i.e. where the IOE approach is not valid. A similar numerical comparison was previously carried out in Ref.~\cite{Andres:2020kfg}. As discussed above, we numerically observe that in the soft sector, $\omega_{\rm BH} \ll \omega \ll \omega_c$, the difference between the full result and the LO contribution is small, and including the NLO provides a very good approximation. In addition, the IOE has no divergences since the matching scale is chosen for each $\omega$ by solving Eq.~\eqref{eq:Q_r}. At frequencies $\omega \gg \omega_c$, we observe that the LO is power suppressed, but the NLO term matches the full result, even faster than the GLV approximation. Overall, the agreement between the LO+NLO result and the full numerical solution is outstanding.

\section{The medium-induced radiative kernel with the IOE}\label{sec:IOE_radiation_spectrum}

After having revised the building blocks of the IOE, we proceed to compute the fully differential medium-induced spectrum for a gluon with energy $\omega$ and transverse momentum $\k$. We assume that the emitted gluon is soft, $\omega \ll E$, and collinear, $\theta^2\sim k_\perp^2/\omega^2\ll 1$, with $E$ being the energy of the emitter. The emitter follows an eikonal trajectory and its kinematics are frozen. Regarding the medium properties, which are encapsulated by the jet quenching parameter $\hat{q}$, we assume that it has a smooth time profile \textit{almost everywhere} and that, at large distances, the system reaches the vacuum sufficiently fast, i.e. $\lim\limits_{t\to \infty}\hat{q}(t)= 0$. Then, we study a particular scenario where the medium has a simple time dependence: up to a distance $L$ the jet quenching parameter is positive and constant, while for times larger than $L$, $\hat{q}\!=\!0$. This corresponds to the previously mentioned plasma brick model, where the medium is a slab with longitudinal size $L$, after which there is vacuum; mathematically it corresponds to defining the jet quenching parameter as $\hat{q}(t)
\!=\! \hat{q}\,\Theta(L-t) $.

Under these assumptions, the \textit{purely} medium-induced spectrum can be expressed as a convolution between the broadening probability distribution, $ \cP$, and the splitting kernel, $\cK$, that we have introduced in the previous section. It reads,
\begin{align}\label{eq:spectrum}
  (2\pi)^2\omega\frac{\rmd I}{\rmd \omega  \rmd^2 \k}&=\lim_{\epsilon\to 0}\frac{2\bar{\alpha}\pi}{\omega^2} \Re \int_0^\infty \rmd t_2 \, \rme^{- \epsilon (t_2+t_1)} \int_0^{t_2}  \rmd t_1 \int_\x \,  \rme^{-i \k \cdot \x}\, \cP(\x,\infty;t_2) \nn
  & \bdel_\x\cdot \bdel_\y \cK(\x,t_2;\y,t_1)_{\y=0}  - (2\pi)^2\omega\frac{ \rmd I^{\rm vac}}{\rmd\omega \rmd^2\k} \, ,
\end{align}
 where $t_1$ and $t_2$ correspond to the gluon splitting light-cone times in amplitude and conjugate amplitude respectively, and span from the creation point inside the medium at $t_1\!=\!t_2\!=\!0$ up to any possible in-vacuum or in-medium splitting time. In \eqn{eq:spectrum}, we explicitly denote the starting time $t_2$ in the broadening distribution so, compared to our formulas in Section~\ref{sec:momentum-broadening}, $\cP(\x,L) \equiv \cP(\x,L;0)$.
Also, in \eqn{eq:spectrum} we employ the adiabatic turn-off prescription~\cite{Wiedemann}, which prevents the emission of purely vacuum-like radiation at asymptotically large times, with the $\epsilon\to 0$ limit being implicit for the rest of the paper. The last term in the formula subtracts a contribution corresponding to purely vacuum radiation off the hard emitter given by (see Appendix~\ref{app:vacuum-derivation})
\beq
\label{eq:spectrum-vac}
 (2\pi)^2\omega\frac{\rmd I^{\rm vac}}{ \rmd \omega  \rmd^2 \k} = \displaystyle\frac{4\bar{\alpha}\pi}{k_\perp^2}.
\eeq
 
Before proceeding further with the explicit analytic evaluation of \eqn{eq:spectrum}, we anticipate a subtlety when carrying out the time integrals with the adiabatic turn-off prescription. Ignoring the $\rme^{-\epsilon t_1}$ suppression factor, the $t_1$ integral can be performed using \eqn{eq:id_1}. Keeping the prescription yields only an additional negative vacuum-like term,  $-(2\pi)^2\omega\frac{ \rmd I^{\rm vac}}{\rmd\omega \rmd^2\k}$, so that \eqn{eq:spectrum} can be expressed in a more convenient form as follows 
\begin{align}\label{eq:spectrum-2}
  (2\pi)^2\omega\frac{\rmd I}{\rmd \omega  \rmd^2 \k}&=\frac{2\bar{\alpha}\pi}{\omega^2} \Re \int_0^\infty \rmd t_2 \, \rme^{- \epsilon t_2} \int_0^{t_2}  \rmd t_1 \int_\x \,  \rme^{-i \k \cdot \x}\, \cP(\x,\infty;t_2) \nn
  & \bdel_\x\cdot \bdel_\y \cK(\x,t_2;\y,t_1)_{\y=0}  - \frac{8\bar{\alpha}\pi}{k_\perp^2} \,, 
\end{align}
where now the $\epsilon$ for the $t_1$ integral prescription has been removed at the cost of a factor 2 multiplying the vacuum term. The limit $\epsilon \to 0$ has to be taken \emph{after} the integral over $t_2$. 
The details regarding the treatment of the adiabatic prescription are discussed in Appendix~\ref{app:vacuum-derivation}.

In what follows, we will compute Eq.~\eqref{eq:spectrum} in the IOE approach, including all terms up to $\mathcal{O}(\delta v)$ (NLO). For that, we extend \eqn{eq:rate-position} for a generic medium and express the broadening distribution $\cP$ as
\beq\label{eq:p-expansion}
 \cP(\x,t;t_0) = \rme^{-\int_{t_0}^t ds\,  \vLO(\x,s) }\rme^{-\int_{t_0}^t ds\, \delta  v(\x,s) }=\cP^{\rm LO}(\x,t;t_0) \, \rme^{-\int_{t_0}^t ds\, \delta  v(\x,s) } \,.
\eeq
Similarly, the emission kernel $\cK$ can be expanded as in \eqn{eq:cK_ful_IOE}:
\begin{align}
\label{eq:k-expansion}
\cK(\x,t_2;\y,t_1)  = \cK^{\rm LO}(\x,t_2;\y,t_1)-\int_\z \int_{t_1}^{t_2} \dd s \, \cK^{\rm LO}(\x,t_2;\z,s)\delta v(\z,s)\cK(\z,s;\y,t_1) \,.
\end{align}
Truncating these relations up to NLO accuracy and inserting them into Eq.~\eqref{eq:spectrum}, we obtain the spectrum which we write in the following way,
\beq
\label{eq:spectrum-ioe}
\frac{\rmd I}{ \rmd \omega  \rmd^2 \k} = \frac{\rmd I^{\rm LO}}{ \rmd \omega  \rmd^2 \k} + \frac{\rmd I^{\rm NLO}}{ \rmd \omega  \rmd^2 \k} + \mathcal{O}(\delta v^2) \,.
\eeq
 To reiterate, the LO and NLO terms resum arbitrary number of \emph{soft} medium interactions, encoded in $\vLO$, and a fixed number (zero for LO, and one for NLO) number of \emph{hard} interactions with the medium, through the potential $\delta v$. 

While the vacuum spectrum is already given in Eq~\eqref{eq:spectrum-vac}, the medium read as follows,
\begin{align}
\label{eq:spectrum-ioe-lo}
(2\pi)^2\omega\frac{\rmd I^{\rm LO}}{ \rmd \omega  \rmd^2 \k}& = \frac{2\bar{\alpha}\pi}{\omega^2} \Re \int_0^\infty \rmd t_2\, \rme^{-\epsilon t_2} \int_0^{t_2} \rmd t_1 \int_\x\, \rme^{-i \k \cdot \x}  \nn
& \times \cP^{\rm LO}(\x,\infty;t_2)  \bdel_\x\cdot \bdel_\y \cK^{\rm {LO}}(\x,t_2;\y,t_1)_{\y=0} - \frac{8 \bar \alpha \pi}{k_\perp^2} \,, \\
\label{eq:spectrum-ioe-nlo}
(2\pi)^2\omega\frac{\rmd I^{\rm NLO}}{ \rmd \omega  \rmd^2 \k}& = \frac{2\bar{\alpha}\pi}{\omega^2} \Re \int_0^\infty \rmd t_2\, \rme^{-\epsilon t_2} \int_0^{t_2} \rmd t_1 \int_\x\, \rme^{-i \k \cdot \x}  \nn
&\times \Big[ \cP^{\rm LO}(\x,\infty;t_2) \bdel_\x\cdot \bdel_\y \cK^{\rm {NLO}}(\x,t_2;\y,t_1)_{\y=0}  \nn
&+ \cP^{\rm NLO}(\x,\infty;t_2) \bdel_\x\cdot \bdel_\y \cK^{\rm {LO}}(\x,t_2;\y,t_1)_{\y=0} \Big]\,,
\end{align}
where
\begin{align}\label{eq:mmmm_1}
\cP^{\rm NLO}(\x, \infty;t) &= - \cP^{\rm LO}(\x,\infty;t) \int_t^\infty \rmd s \, \delta v(\x,s) \, ,
\end{align}
and
\begin{align}\label{eq:mmmm_2}
\cK^{\rm NLO}(\x,t_2;\y,t_1) &= - \int_\z \int_{t_1}^{t_2} \rmd s\, \cK^{\rm  LO}(\x,t_2;\z,s) \delta v(\z,s) \cK^{\rm LO}(\z,s;\y,t_1) \,.
\end{align}
The LO term captures the physics associated with the production of gluon radiation due to multiple soft scattering in the medium, thus recovering the BDMPS-Z solution. The first term in the NLO term \eqn{eq:spectrum-ioe-nlo} includes the possibility of producing the gluon due to a hard scattering in the medium and when integrated over $\k$ gives the NLO contribution to the integrated spectrum studied in the previous section, Eq.~\eqref{eq:dIdw_LO_p_NLO} \cite{IOE1}. Finally, the last term in  \eqn{eq:spectrum-ioe-nlo} arises from expanding the final state broadening distribution $\cP$. Thus, it only affects the redistribution of the radiated gluon transverse momentum and it vanishes upon integration over $\k$. 
In the following sections, we proceed to explicitly compute Eqs.~\eqref{eq:spectrum-ioe-lo} and \eqref{eq:spectrum-ioe-nlo}.

\subsection{Leading order contribution}\label{sec:LO}

The leading order contribution to the spectrum is captured by Eq.~\eqref{eq:spectrum-ioe-lo}. The broadening distribution $\cP^{\rm LO}$, implicitly given in \eqn{eq:p-expansion}, reads
\begin{align} \label{eq:P_LO_x}
  \cP^{\rm LO}(\x,t;t_0) = \exp\left[ -\frac{1}{4}Q_{s0}^2(t,t_0) \log \frac{Q_b^2}{\mu_\ast^2}\, x_\perp^2 \right]\, ,
 \end{align}
 where we define the \textit{bare} saturation scale as a slight generalization of \eqn{eq:old_Qs0}, reading
 \begin{align}
 Q_{s0}^2(t,t_0) = \int_{t_0}^{t}  \rmd s\, \hat q_0(s) \, ,
\end{align}
and the matching scale, $Q_b$, satisfies (following \eqn{eq:old_Qb})
\begin{align}\label{eq:Qb_new}
  Q_b^2\equiv  \int_0^{ \infty}  \rmd t\, \hat q_0(t)   \,\log \frac{Q_b^2}{\mu_\ast^2}\, .
 \end{align}
 Furthermore, the kernel $\cK^{ \rm LO}$ can be found in Eq.~\eqref{eq:cK_BDMPS_2} but, for consistency, we also repeat it here in a slightly different form, namely
  \begin{align}\label{eq:cK_BDMPS_3}
   \cK^{\rm LO}(\x,t_2;\y,t_1)  = \frac{\omega}{2\pi i S(t_2,t_1)} \exp \left[i\frac{\omega}{2} \left( {\rm Cot}(t_2,t_1) \, \x^2 - {\rm Cot}(t_1,t_2) \, \y^2 - \frac{2}{S(t_2,t_1) }\, \x\cdot \y \right) \right] \,,
  \end{align}
where we recall that (see Appendix~\ref{app:cK_appendix} for further details)
\beq
\label{eq:cot}
{\rm Cot}(t_2,t_1) = \frac{C(t_1,t_2)}{S(t_2,t_1)} \,.
\eeq
Also, in \eqn{eq:cK_BDMPS_3} we have taken advantage of the anti-symmetry of $S$, i.e. $S(t_2,t_1) = - S(t_1,t_2)$.
  
In all these functions, the value of $\hat q$ enters as an argument and thus they are sensitive to the definition of the matching scale for radiation, $Q_r$, that, as discussed in Section~\ref{sec:gluon_generic_IOE}, is obtained by solving the transcendental equation (see \eqn{eq:Q_r})
  \beq
  \label{eq:matching-q2r}
  Q^2_r(t)=\sqrt{\hat q(t) \omega}=\sqrt{\hat q_0(t) \omega  \log \frac{Q^2_r(t)}{\mu_\ast^2}}.
    \eeq  
 At this point, an important remark is in order. The functional form of the matching scale is constrained by making the spectrum finite in the infrared. This leads to the $\omega$-dependence in Eq.~\eqref{eq:matching-q2r}, which is constrained in this way up to an overall numerical coefficient. As was shown in Ref.~\cite{IOE3}, the dependence in such a factor is sub-leading for a fixed order calculation in the IOE. As such, and since all the time dependence of $Q_r^2$ emerges from $\hat{q}_0$, it is more convenient to  define $Q_r^2$ as a static scale. This simplifies the time integrations needed for the spectrum calculation without downgrading its accuracy.
   
In order to further simplify Eq.~\eqref{eq:spectrum-ioe-lo}, we make use of a series of identities satisfied by the functions $C(t_2,t_1)$ and $S(t_2,t_1)$ that enter in the definition of $\cK^{ \rm LO}$, see \eqn{eq:cs-ho-equations}. In particular, we use Eq.~\eqref{eq:prop_app_1} to obtain
  \begin{align}\label{eq:id_1}
\int_0^{t_2} \rmd t_1 \, \partial_\y \cK^{\rm LO}(\x,t_2;\y,t_1)_{\y=\0} &= -\frac{\omega^2}{2\pi}\int_0^{t_2} \rmd t_1\, \frac{\x}{S^2(t_2,t_1)}\rme^{\frac{i\omega}{2} {\rm Cot}(t_2,t_1) \x^2} \nn 
&= \frac{\omega}{\pi i} \frac{\x}{\x^2} \rme^{\frac{i\omega }{2} {\rm Cot}( t_2,0) \x^2} \, ,
\end{align} 
where in the second step we dropped an infinite phase which has already been accounted for in the vacuum subtraction term in \eqn{eq:spectrum-2}. A careful treatment of this technical point is presented in Appendix~\ref{app:vacuum-derivation}, see \eqn{eq:t1-int} and related discussion. 
The identity \eqn{eq:id_1} together with Eq.~\eqref{eq:cK_BDMPS_2} lead to the following expression for the leading-order spectrum  
\begin{align}
\label{eq:HO_any_med_1a}
(2\pi)^2\omega\frac{\rmd I^{\rm LO}}{\rmd \omega \rmd^2 \k} &=2\bar{\alpha} \Re \int_0^\infty  \rmd t_2 \, \rme^{-\epsilon t_2}\,  {\rm Cot}(t_2,0)\int_\x \rme^{-i \k \cdot \x} \, \cP^{\rm LO}(\x,\infty;t_2) \,  \rme^{\frac{i\omega }{2} {\rm Cot}(t_2,0)\, \x^2}  \nn
&-\frac{8\pi \abar}{k_\perp^2} \, ,
\end{align}
where the term proportional to $\partial_\x \cdot \frac{\x}{\x^2}\propto \delta^{(2)}(\x)$ is purely imaginary and thus does not contribute given the $\epsilon$ prescription adopted in \eqn{eq:spectrum-2} (see Appendix~\ref{app:vacuum-derivation} for more details).

The result obtained in Eq.~\eqref{eq:HO_any_med_1a}, although compact, is somewhat obscure from a physical perspective. A more intuitive description of the in-medium emission can be achieved by using a momentum space representation
\begin{align}\label{eq:HO_any_med_2a}
(2\pi)^2\omega\frac{\rmd I^{\rm LO}}{\rmd \omega \rmd^2 \k}
&=\frac{4\bar{\alpha}\pi}{\omega} \Re\, i\int_0^\infty \rmd t_2 \, \rme^{-\epsilon t_2}\,
\int_{\p}  \, \cP^{\rm LO}(\k-\p,\infty;t_2)  \rme^{-i\frac{\p^2 }{2 \omega {\rm Cot}(t_2,0)}  }-\frac{8\pi \abar}{k_\perp^2}  \, .
\end{align}
In this form, we can interpret the first term in Eq.~\eqref{eq:HO_any_med_2a} as describing the emission of a gluon via some effective kernel at time $t_2$ followed by final state broadening. The second term corresponds to a vacuum-like subtraction contribution.

Furthermore, since $\cP^{\rm LO}(\p)$ is Gaussian, the remaining momentum integral can be performed 
\beq
\int_{\p}  \, \cP^{\rm LO}(\k-\p,\infty;t) \,  \rme^{- i\frac{\p^2 }{2\omega  {\rm Cot}(t,0)}  }  =  -2i\omega  {\rm Cot}(t,0)\, \frac{ \rme^{-\frac{\k^2}{\khat^2(t,0)} }}{\khat^2(t,0)} \, , 
\eeq
where we introduced the function
\beq
\label{eq:Lambda21}
\khat^2(t_2,t_1) = Q^2_s(\infty,t_2) -2i\omega{\rm Cot}(t_2,t_1)\, ,
\eeq
and the \textit{effective} saturation scale is defined as $Q_s^2(t_2,t_1) = \int_{t_1}^{t_2} \rmd t \,  \hat{q}_0(t) \log \frac{Q_b^2}{\mu_\ast^2}$, with the logarithmic dependence in $\hat{q}$ determined by $Q_b$. Inserting this result into Eq.~\eqref{eq:HO_any_med_2a}, we finally obtain
\begin{align}\label{eq:HO_any_med_3}
(2\pi)^2\omega\frac{\rmd I^{\rm LO}}{\rmd \omega \rmd^2 \k}=8\Bar{\alpha} \pi \, \Re  \int_0^\infty \rmd t  \,  \rme^{-\epsilon t}\,  \frac{{\rm Cot}(t,0)}{\khat^2(t,0)}  \rme^{-\frac{k_\perp^2}{\khat^2(t,0)}}  -\frac{8\pi \abar}{k_\perp^2} \,.
\end{align}
This form of the spectrum was first derived in Ref.~\cite{Caucal:2020zcz}. Integrating over $\k$ and using  $\int_\k \cP(\k)=1$, we recover the LO contribution to the energy spectrum discussed in the previous section \cite{IOE1,Arnold_simple}, see Eq.~\eqref{eq:dIdw_BDMPS}. 

As a sanity check, one can verify that \eqn{eq:HO_any_med_3} vanishes in the vacuum, i.e. in the limit $\Omega \to 0$, so that
\beq
{\rm Cot}(t,0) \to \frac{1}{t} \, \quad \text{and} \quad \khat^2(t_2,t_1) \to -\frac{2i\omega}{ t_2-t_1}\,.
\eeq
Thus, we obtain 
\begin{align}\label{eq:HO_any_vac_3}
(2\pi)^2\omega\frac{\rmd I^{\rm LO}}{\rmd \omega \rmd^2 \k}=8\Bar{\alpha} \pi \, \Re  \int_0^\infty \rmd t  \,  \rme^{-\epsilon t} \,\frac{1}{2i \omega }  \rme^{-i \frac{k_\perp^2}{2\omega}t}  -\frac{8\pi \abar}{k_\perp^2}=0 \,.
\end{align}
\paragraph*{Plasma brick model.} 
We proceed to evaluate the previous expressions for a concrete medium model, namely the brick where $\hat{q}(t)= \hat{q}\, \Theta(L-t)$. Inside the medium, i.e. when both $t_1<L$ and $t_2<L$, the $C$ and $S$ functions take simple forms (see Appendix~\ref{app:cK_appendix})
\begin{align}\label{eq:SC-brick}
S(t_2,t_1)=\frac{\sin(\Omega(t_2-t_1))}{\Omega} \, , \quad C(t_2,t_1)=\cos(\Omega(t_2-t_1))  \, ,
\end{align}
and ${\rm Cot}(t_2,t_1) = \Omega \cot (\Omega (t_2-t_1))$, where $\Omega=\frac{1-i}{2}\sqrt{\frac{\hat{q}_0}{ \omega} \log \frac{Q_r^2}{\mu_\ast^2}}$.
 On the other hand, for both $t_1>L$ and $t_2 >L$, the system evolves as in vacuum and the $C$ and $S$ are obtained by setting $\Omega\to 0$ in the previous equations such that 
\begin{align}\label{eq:SC-vacuum}
S(t_2,t_1) = t_2-t_1 \, , \quad C(t_2,t_1)=1 \,. 
\end{align}
This sharp separation of the problem into processes happening inside the medium and outside of it, suggests that an efficient way to evaluate Eq.~\eqref{eq:HO_any_med_3} consists in splitting the time integral into two regions: one where $0<t<L$ and a vacuum-like region where $t>L$. 

We refer to the first region as \textbf{in-in} since the gluon emission occurs inside the medium in both the amplitude and its conjugate. It can be easily obtained by replacing the upper limit in the integral in Eq.~\eqref{eq:HO_any_med_3} by $L$ and employing Eq.~\eqref{eq:SC-brick}. Since the integral over $t_2$ has a finite extension, we can safely neglect the adiabatic suppression factor. This contribution to the total medium-induced LO spectrum reads
\begin{align}\label{eq:HO_brick_InIn}
(2\pi)^2\omega\frac{\rmd I^{\rm LO}_{\text{in-in}}}{\rmd \omega \rmd^2 \k}&=8\Bar{\alpha} \pi \, \Re \int_0^L \rmd t  \, \Omega {\rm cot}(\Omega t)   \frac{ \rme^{-\frac{k_\perp^2}{\hat{q}(L-t) -2i  \omega  \Omega \cot(\Omega t)} }}{\hat{q}(L-t) - 2i  \omega \Omega {\rm cot}(\Omega t)}\, ,
\end{align}
where we have used that the saturation scale reduces to $Q_s^2(\infty,t)=\hat{q}(L-t)$.

The remaining region of phase space is obtained by imposing $t>L$ in Eq.~\eqref{eq:HO_any_med_3}. In this case, there are two types of contributions: (i) a purely vacuum term corresponding to the scenario where the gluon is outside the medium both in the amplitude and its conjugate, in this situation the first and second terms in Eq.~\eqref{eq:HO_any_med_3} cancel by construction, and (ii) an interference term where the amplitude gluon is emitted inside the medium while its conjugate counterpart is emitted in the vacuum (or vice-versa). The latter contribution, which we shall refer to as \textbf{in-out}, requires further manipulations. 

We begin by constructing the $C$ and $S$ functions which have support inside and outside the medium. This is done by using the decomposition for the $C$ and $S$, given in Eq.~\eqref{eq:linear_rel_C_S}~\cite{Arnold_simple} 
\begin{equation}
  \begin{split}
  &S(t_2,t_1)=C(t_1,t_0)S(t_2,t_0)-S(t_1,t_0)C(t_2,t_0) \, ,\\ 
 & C(t_2,t_1)=-\partial_{t_1}C(t_1,t_0)S(t_2,t_0)+\partial_{t_1}S(t_1,t_0)C(t_2,t_0) \, .
  \end{split}    
\end{equation}
Taking $t_1=0$, $t_2=t>L$, $t_0=L$ and using the appropriate form for the $C$ and $S$ in each region (see Eqs.~\eqref{eq:SC-brick} and \eqref{eq:SC-vacuum}), we obtain
\begin{equation}
  \begin{split}
&S(t,0)=(t-L) \cos \Omega L + \frac{\sin \Omega L}{\Omega }\, ,\\
&C(0,t)=\cos(\Omega L)\, , 
\end{split}    
\end{equation}
which yields
\beq
 {\rm Cot} (t,0)= \frac{ \Omega \cot\Omega L}{\Omega (t-L) \cot \Omega L +1 }\, .
\eeq
In addition, in Eq.~\eqref{eq:SC-brick} the broadening term (encapsulated in $Q_s^2$) has support in $(t,L)$ which is the vacuum region. Then, there is no final state broadening and one can take
\beq
\cP^{\rm LO}(\k-\p,\infty;t) \big|_{t>L}=   (2\pi)^2 \delta^{(2)}(\k) \, ,
\eeq
or, equivalently, $Q_s^2=0$ in  Eq.~\eqref{eq:SC-brick}. Combining all these results, we find that 
  \begin{align}\label{eq:HO_brick_out_2}
(2\pi)^2\omega\frac{\rmd I_{\text{in-out}}^{\rm LO}}{\rmd \omega \rmd^2 \k}
&=\frac{4\bar{\alpha}\pi}{\omega} \Re \,i  \int_L^\infty \rmd t
\,  \rme^{-i\frac{k_\perp^2  }{2\omega  }(t-L)- i \frac{k_\perp^2}{2\omega \Omega \,  \cot\Omega L } } -\frac{8\pi \abar}{k_\perp^2} \, \nn
&= \frac{8\bar{\alpha}\pi }{k_\perp^2}\Re\left(\rme^{-i\frac{k_\perp^2}{2\omega \Omega \,  \cot\Omega L }  }- 1\right) \, ,
\end{align}
where we have included the vacuum subtraction term, and where the implicit adiabatic prescription $\sim\rme^{-\epsilon t}$  in the first line allowed to drop the contribution from $t\to \infty$.

This contribution, together with Eq.~\eqref{eq:HO_brick_InIn}, constitute the medium-induced leading order spectrum, analogous to the BDMPS-Z result. The medium-induced spectrum at LO in the IOE reads then,
\beq
\frac{\rmd I^{\rm LO}}{\rmd \omega \rmd^2 \k} = \frac{\rmd I^{\rm LO}_{\text{in-in}}}{\rmd \omega \rmd^2 \k} + \frac{\rmd I^{\rm LO}_{\text{in-out}}}{\rmd \omega \rmd^2 \k} \,.
\eeq

\subsection{Next-to-leading order contribution}\label{sec:nlo}

The computation of the next-to-leading order contribution to the spectrum can be done using similar manipulations to the ones performed for the LO term in the previous section. The NLO spectrum is defined in Eq.~\eqref{eq:spectrum-ioe-nlo}. The first term corresponds to a genuine correction to the emission kernel, referred below to as the \textbf{in} contribution, while the second term, which we shall refer to as \textbf{broad} contribution, introduces the possibility of a hard scattering in the final state broadening process. Also, the vacuum-like subtraction terms that appeared in the LO contribution are absent at $\mathcal{O}(\delta v)$. 

To summarize, the two contributions to the NLO spectrum read 
\begin{align}
\label{eq:spectrum-ioe-nlo-in}
  (2\pi)^2\omega\frac{\rmd I^{\rm NLO}_{\rm in}}{\rmd \omega \rmd^2 \k} &= -\frac{2\bar{\alpha}\pi}{\omega^2} \Re \int_0^\infty \rmd t_2 \int_0^{t_2} \rmd t_1 \int_0^{t_1} \rmd s \int_{\x,\z}\,  \rme^{-i \k \cdot \x} \, \cP^{\rm LO}(\x,\infty ;t_2)   \nn
& \times \delta v(\z,t_1) \bdel_\x   \cK^{\rm {LO}}(\x,t_2;\z,t_1) \cdot \bdel_\y \cK^{\rm {LO}}(\z,t_1;\y,s) \,,\\
\label{eq:spectrum-ioe-nlo-broad}
(2\pi)^2\omega\frac{\rmd I^{\rm NLO}_{\rm broad}}{\rmd \omega \rmd^2 \k} &= -\frac{2\bar{\alpha}\pi}{\omega^2} \Re \int_0^\infty \rmd s \int_0^s \rmd t_2 \int_0^{t_2} \rmd t_1 \int_\x  \rme^{-i \k \cdot \x} \, \cP^{\rm LO}(\x,\infty,t_2)   \nn
& \times \delta v(\x,s)  \bdel_\y \cdot \bdel_\x \cK^{\rm LO}(\x,t_2;\y,t_1) \bigg)_{\y=0} \,,
\end{align}
where we have rearranged the time integrations. \footnote{Note that in \eqn{eq:spectrum-ioe-nlo-in} $t_1$ is the time at which the hard interaction occurs, while we have labelled this time as $s$ in other equations.} 
Let us begin by considering  Eq.~\eqref{eq:spectrum-ioe-nlo-in}. Note, that both the $t_1$ and $s$ integrals have support only inside the medium, hence the naming as the \textbf{in} contribution. By using Eq.~\eqref{eq:id_1}, we can directly perform the $s$-integral  such that
\begin{align}\label{eq:IOE_med_2}
(2\pi)^2\omega\frac{ \rmd I_{\text{in}}^{\rm NLO}}{\rmd\omega \rmd^2 \k}&=\frac{2\bar{\alpha}}{\omega} \,\Re\, i \int_0^\infty \rmd t_2 \int_\x \,\rme^{-i \k \cdot \x} \, \cP^{\rm LO}(\x,\infty; t_2)  
\nn
&\times \int_0^{t_2} \rmd t_1  \int_\z \,\delta v(\z,t_1) \frac{\z}{\z^2}\cdot \partial_\x \cK^{\rm LO}(\x,t_2;\z,t_1) \rme^{\frac{i\omega}{2}{\rm Cot}(t_1,0)\z^2}
 \, .
\end{align}
The remaining derivative operator gives 
\begin{align}
  \partial_\x \cK^{\rm LO}(\x,t_2;\z,t_1) &= \frac{\omega^2}{2\pi S^2(t_2,t_1)}  \big(\x C(t_1,t_2)-\z \big) \nn 
 &\times \exp\left[\frac{i\omega}{2S(t_2,t_1)} \big( C(t_1,t_2) \x^2 + C(t_2,t_1)\z^2 - 2 \x\cdot\z \big)\right] \, ,
  \end{align}
so that the spectrum further simplifies to (adopting hereafter the more compact notation ${\rm Cot}_{21} \equiv {\rm Cot}(t_2,t_1)$, $C_{12}\equiv C(t_1,t_2)$ and $S_{12}\equiv S(t_2,t_1)$) 
\begin{align}
\label{eq:IOE_med_3}
(2\pi)^2\omega\frac{\rmd I_{\text{in}}^{\rm NLO}}{\rmd \omega \rmd^2 \k} & =\frac{\bar{\alpha}\omega}{\pi}\Re\, i \int_0^\infty \rmd t_2\int_0^{t_2} \rmd t_1 \int_{\x,\z}\, \rme^{-i \k \cdot \x}\rme^{-\frac{1}{4}Q_s^2(\infty,t_2) \x^2} \nn
&\times \frac{1}{S^2_{21}} \delta v(\z,t_1) \frac{\z}{\z^2}\cdot (\x C_{12}-\z)
\nn 
&\times \rme^{\frac{i\omega}{2S_{21}} \left(C_{12}\x^2+C_{21}\z^2-2\x\cdot\z\right)} \rme^{\frac{i\omega}{2}{\rm Cot}_{10} \z^2}  \, .
\end{align}
The remaining integration in $\x$ is Gaussian, and can be executed to obtain
\begin{align}
  \int_\x \rme^{-i \k \cdot \x}\rme^{-\frac{1}{4}Q_s^2(\infty,t_2) \x^2} & \rme^{\frac{i\omega}{2S_{21}} \left(C_{12}\x^2 - 2\x\cdot\z \right)}(\x C_{12}-\z)
  \nn 
  =&-\frac{4\pi}{\khat^2_{21}} \rme^{-\frac{\left(\k-\frac{\omega }{S_{12}} \z\right)^2}{\khat^2_{21}}}\left[\z+\frac{2iC_{12}}{\khat^2_{21}}\left(\k-\frac{\omega}{S_{12}} \z\right)\right] \, .
\end{align}
Let us re-emphasize that, in the previous expression, the matching scale associated with $Q_s$ in $\khat_{21}$ is $Q_b$. Replacing $\delta v$ by its explicit definition, that depends on $Q_r$, the $\textbf{in}$ contribution reads
\begin{align}
\label{eq:IOE_med_4}
&(2\pi)^2\omega\frac{\rmd I_\text{in}^{\rm NLO}}{\rmd \omega \rmd^2 \k} =-\bar{\alpha}\omega \Re\, i\int_0^\infty \rmd t_2 \int_0^{t_2}  \rmd t_1 \,  \hat{q}_0(t_1) \frac{\rme^{-\frac{k_\perp^2}{\khat^2_{21}}}}{S^2_{21}\khat^4_{21}} \nn
&\times \int_\z \, \rme^{\frac{i\omega}{2}(-{\rm Cot}_{12}+{\rm Cot}_{10}+\frac{2i\omega}{S^2_{21}\khat^2_{21}}) \z^2} \rme^{-\frac{2\omega}{\khat^2_{21}S_{21}} \k\cdot \z}
 \log \frac{1}{Q_r^2 \z^2} \left(Q^2_s(\infty,t_2) \z^2+2iC_{12}\k\cdot \z\right)  \, .
\end{align}
Notice that we tend to write the largest time as the first argument in the functions. Nonetheless, this is not always possible since in general the $C$ function has no definite parity under the exchange of the arguments, unlike the $S$ function which is always odd. 

Comparing Eq.~\eqref{eq:IOE_med_4} to the LO contribution given by Eq.~\eqref{eq:HO_any_med_1a}, we observe an additional transverse integral in the $\z$ variable which is no longer Gaussian due to the logarithmic dependence in $\delta v$. Nevertheless, this integration can be performed analytically too. The angular part can be performed by recalling the definitions of the Bessel functions of the first kind, 
\begin{align}
  \int_0^{2\pi} \frac{\rmd \theta}{2\pi } \rme^{-i z \cos\theta}  =  J_0(z)    \, , \quad \int_0^{2\pi} \frac{\rmd \theta}{2\pi } \cos\theta \, \rme^{-i z \cos\theta}   = -i J_1(z)   \,, 
\end{align}
that lead to 
\begin{align}
\label{eq:nlo-in}
&(2\pi)^2\omega\frac{\rmd I_\text{in}^{\rm NLO}}{\rmd \omega \rmd^2 \k}=\frac{\bar{\alpha}\pi}{2\omega k_\perp^4} \Re \, i\int_0^\infty \rmd t_2 \,\int_0^{t_2}  \rmd t_1 \,  \frac{\hat{q}_0(t_1)}{\rhat_{21}^2} \rme^{-\frac{k_\perp^2}{\khat^2_{21}}} 
\nn
&\times \int_0^\infty  \rmd z_\perp \,  z_\perp \,  \rme^{-\jhat_{21}\frac{z_\perp^2}{4 k_\perp^2} } \log \left(\frac{k_\perp^2\rhat^2_{21}}{Q_r^2z_\perp^2}\right) \left[Q^2_s(\infty,t_2) z_\perp^2J_0(z_\perp)+2 C_{12}\rhat_{21} k_\perp^2 z_\perp J_1( z_\perp) \right]  \, ,
\end{align}
where we have introduced the auxiliary functions
\begin{align}
 &\jhat(t_2,t_1)\equiv \jhat_{21}=\frac{i}{2\omega}\left(-{\rm Cot}_{12}+{\rm Cot}_{10}+\frac{2i\omega}{S^2_{21}\khat^2_{21}}  \right)S_{21}^2 \khat_{21}^4\, , \nn 
 &\rhat(t_2,t_1)\equiv\rhat_{21} =-\frac{2i\omega}{\khat^2_{21}S_{21}} \, .
\end{align}
The more challenging radial integral can be solved by using a convenient decomposition of the logarithmic function. Namely, the relation
\begin{equation}\label{eq:Moliere_log}
  \log\frac{1}{u^2}=\lim_{\epsilon\to0} \int_\epsilon^\infty \frac{\rmd t}{t}    \, \left(\rme^{-u^2t}-\rme^{-t}\right) \, ,
  \end{equation}
allows us to transform the original integral into a sum of Gaussian integrations that can be readily performed. In particular, the $z$-integrals in Eq.~\eqref{eq:nlo-in} can be compactly expressed as
\beq
I_a(x,y ) =  \int_0^\infty \rmd z J_0(z)\, z^3 \,\log\frac{y}{z^2}   \, \rme^{ - \frac{z^2}{4x}}\, ,
\eeq
and 
\begin{align}
I_b(x,y )=\int_0^\infty \rmd z \, z^2 \log \frac{y}{z^2} J_1\left(z \right)\rme^{-\frac{z^2}{4x}} \, .
\end{align}
Then, replacing the logarithms in the previous equations by the decomposition in Eq.~\eqref{eq:Moliere_log} allows one to write $I_a$ and $I_b$ in terms of the exponential integral function $\rm Ei$, leading to 
\beq\label{eq:bbS_alpha}
I_a\left(x,y \right) 
&=&8 x^2\, \rme^{-x}(-2+\rme^{x})  + 8 x^2 \, \rme^{-x}(1-x)\left[{\rm Ei}\left(x\right)-\log\frac{4x^2 }{y}\right]\, ,
\eeq
and
\begin{align}\label{eq:bbS_beta}
  &I_b\left(x,y\right)=-4x\left(1-\rme^{-x}\right)+4x^2\rme^{-x}\left[{\rm Ei}(x)-\log \frac{4x^2}{y}\right]\, .
  \end{align}
Taking advantage of these further simplifications, the $\mathbf{in}$ contribution to the NLO gluon spectrum can be compactly written as 
\begin{align}\label{eq:IOE_NLO_IN}
(2\pi)^2\omega\frac{\rmd I_{\text{in}}^{\rm NLO}}{\rmd \omega  \rmd^2 \k}&=\frac{\bar{\alpha}\pi}{2\omega k_\perp^4} \Re\, i\int_0^\infty \rmd t_2 \int_0^{t_2} \rmd t_1 \,  \frac{\hat{q}_0(t_1)}{\rhat_{21}^2} \rme^{-\frac{ k_\perp^2}{\khat^2_{21}}} 
\nn
&\times  \left[Q^2_s(\infty,t_2) I_a\left(\frac{k_\perp^2}{\jhat_{21}},\frac{k_\perp^2\rhat^2_{21}}{Q_r^2}\right)+2 C_{12}\rhat_{21} k_\perp^2 I_b\left(\frac{k_\perp^2}{\jhat_{21}},\frac{k_\perp^2\rhat^2_{21}}{Q_r^2}\right)\right] \,.
\end{align}
Hence, we have managed to reduce the number of integrals over transverse positions and times down to two time-integrations.

Turning now to the \textbf{broad} contribution, given in \eqn{eq:spectrum-ioe-nlo-broad}, we can perform the derivatives on $\cK^{\rm LO}$ and integrate over $t_1$ using \eqn{eq:id_1}. Then it reads
\begin{align}\label{eq:IOE_med_broad1}
  (2\pi)^2\omega\frac{
  d I_{\rm \text{broad}}^{\rm NLO}}{\rmd \omega \rmd^2\k}&=-2\bar{\alpha} \Re \int_0^\infty \rmd t_2 \int_{t_2}^\infty \rmd s \int_\x \rme^{-i \k \cdot \x} \, \cP^{\rm LO}(\x,\infty;t_2)\nn 
&\times \delta v(\x,s)\, {\rm Cot}_{20}  \, \rme^{\frac{i\omega}{2}{\rm Cot}_{20}\x^2}  \, ,
  \end{align}
with $\cP^{\rm LO}(\x,\infty;t_2) $ introduced in Eq.~\eqref{eq:P_LO_x}. Using the definition of the  function $I_b(x,y)$ in \eqn{eq:bbS_alpha}, the $\mathbf{broad}$ contribution to the spectrum can finally be written as
\begin{align}\label{eq:IOE_med_broad2}
(2\pi)^2\omega\frac{
  \rmd I_{\rm \text{broad}}^{\rm NLO}}{\rmd \omega \rmd^2 \k}
&= -\frac{\pi \bar{\alpha}}{k_\perp^4} \Re \int_0^\infty \rmd t_2 \, {\rm Cot}_{20}\, Q^2_{s0}(\infty,t_2) \, I_a\left(\frac{k_\perp^2}{\khat^2_{20}},\frac{k_\perp^2}{Q_b^2}\right)\bigg]\, ,
\end{align}
where $\khat^2_{20} \equiv\khat^2(t_2,0)$ given by \eqn{eq:Lambda21}.

\paragraph*{Plasma brick model.} So far, we have made no approximation regarding the time profile of the medium. As for the LO contribution, we now assume the plasma brick model, i.e. $\hat{q}(t)=\hat{q}\,\Theta(L-t)$. As in the previous case, this simple model allows one to further simplify Eqs.~\eqref{eq:IOE_NLO_IN} and \eqref{eq:IOE_med_broad2} by splitting the time integrations appropriately.

We start with the \textbf{broad} term since it has only one time integral left. In addition, the spectrum is proportional to $Q_{s0}^2 $, and thus this term only has support inside the medium $t_2<L$. As a consequence, the $C$ and $S$ functions are given directly by Eq.~\eqref{eq:SC-brick}, and Eq.~\eqref{eq:IOE_med_broad2} reduces to 
\begin{align}\label{eq:IOE_med_broad_brick}
  (2\pi)^2\omega\frac{
    \rmd I_{\rm \text{broad}}^{\rm NLO}}{\rmd \omega \rmd^2 \k}
  &= -\frac{\hat{q}_0\pi \bar{\alpha}}{k_\perp^4} \Re\int_0^L \rmd t_2 \, \Omega \, {\rm cot}(\Omega t_2)\, (L-t_2) \,I_a\left(\frac{k_\perp^2}{\khat^2(t_2,0)},\frac{k_\perp^2}{Q_b^2}\right) \, ,
  \end{align}
with 
\beq\label{eq:help_1}
 \khat^2(t_2,0)= Q^2_s(L,t_2) -2i\omega\Omega{\rm cot}(\Omega t_2)\, , \quad  Q^2_s(L,t_2)=\hat{q}_0(L-t_2) \log \frac{Q_b^2}{\mu_\ast^2} \, .
\eeq
Next, let us analyze the \textbf{in} contribution given by Eq.~\eqref{eq:IOE_NLO_IN}. It has support both inside and outside of the medium and thus two contributions appear. One option is that the gluon is emitted inside the medium in the amplitude and its conjugate, which we identify as the \textbf{in-in} contribution. The second term refers to the case in which one of the emissions happens outside of the medium, that we denote as \textbf{in-out}. The \textbf{in-in} term obeys the time ordering $\int_0^L \rmd t_2 \int_0^{t_2} \rmd t_1$ in Eq.~\eqref{eq:IOE_NLO_IN}, with the $C$ and $S$ having support only inside the medium and thus are given by Eq.~\eqref{eq:SC-brick}. Therefore, this contribution reads
\begin{align}\label{eq:IOE_NLO_ININ}
  (2\pi)^2\omega\frac{\rmd I_{\text{in-in}}^{\rm NLO}}{\rmd \omega \rmd^2 \k}&=\frac{\bar{\alpha}\pi\hat{q}_0}{2\omega k_\perp^4} \Re\, i\int_0^L \rmd t_2 \,\int_0^{t_2} \rmd t_1 \,  \frac{\rme^{-\frac{k_\perp^2}{\khat^2_{21}}}}{\rhat_{21}^2}  
  \nn
  &\times  \left[Q^2_s(L,t_2) I_a\left(\frac{k_\perp^2}{\jhat_{21}},\frac{k_\perp^2\rhat^2_{21}}{Q_r^2}\right)+2 C_{12}\rhat_{21} k_\perp^2 I_b\left(\frac{k_\perp^2}{\jhat_{21}},\frac{k_\perp^2\rhat^2_{21}}{Q_r^2}\right)\right] \,,
  \end{align}
where again $Q^2_s(L,t_2)\!=\!\hat{q}_0(L-t_2)\log \frac{Q_b^2}{\mu_\ast^2} $, and the auxiliary functions reduce to
\begin{align}\label{eq:help_2}
  & \khat^2_{21} = Q^2_s(L,t_2) -2i\omega \Omega \, {\rm cot}(\Omega (t_2-t_1))\, , \nn
  & \jhat_{21}=\frac{i }{2\omega}\left(\Omega{\rm cot}(\Omega(t_2-t_1))+\Omega{\rm cot}(\Omega t_1)+\frac{2i\omega}{S^2_{21}\khat^2_{21}}  \right)S_{21}^2 \khat_{21}^4\, , \nn 
  &\rhat_{21} =-\frac{2i\omega}{\khat^2_{21}S_{21}} \, .
 \end{align}
The \textbf{in-out} contribution can be further simplified following similar steps as in the LO case. Now, the time integrals read $\int_L^{\infty} \rmd t_2 \int_0^{L}\rmd  t_1 $ and one can set $Q_s^2=0$ everywhere in Eq.~\eqref{eq:IOE_NLO_IN} since this term only has support outside of the medium. Then, the spectrum reads
\begin{align}\label{eq:IOE_NLO_INPOUT_1}
&(2\pi)^2\omega\frac{\rmd I_{\text{in-out}}^{\rm NLO}}{\rmd \omega \rmd^2 \k}=\frac{\bar{\alpha}\hat{q}_0\pi}{\omega k_\perp^2} \Re\, i\int_L^\infty \rmd t_2 \,\int_0^{L} \rmd t_1 \,  \frac{C_{12}}{\rhat_{21}} \rme^{-\frac{ k_\perp^2}{\khat^2_{21}}} 
I_b\left(\frac{k_\perp^2}{\jhat_{21}},\frac{k_\perp^2\rhat^2_{21}}{Q_r^2}\right)  \, ,
\end{align}
where we recall that the $C$ and $S$ functions are distinct from the ones used in the \textbf{in-in} term, since they have support both inside and outside of the medium. Nonetheless, as for the LO case, they can be written in terms of the purely in-medium and in-vacuum $C$ and $S$ functions by using Eq.~\eqref{eq:linear_rel_C_S}. Taking $t_0=L$, we find that for the above time ordering 
\begin{align}
 & S_{21}=C_{1L}S_{2L}-S_{1L}C_{2L}=\cos(\Omega (L-t_1))(t_2-L)+\frac{\sin(\Omega(L-t_1))}{\Omega}\, , \nn    
 & C_{12}=-\partial_2 S_{12}=\partial_2 S_{21}=\cos(\Omega(L-t_1)) \, ,
  \end{align}
such that 
\begin{align}
  {\rm Cot}_{21}&=\frac{C_{12}}{S_{21}}=\frac{\Omega}{\Omega (t_2-L) + \tan (\Omega(L-t_1))} \, . 
\end{align}
For the reversed time ordering, we find 
\begin{align}
  S_{12}&=-S_{21}=-\cos(\Omega (L-t_1))(t_2-L)-\frac{\sin(\Omega(L-t_1))}{\Omega} \, , \nn    
  C_{21}&=-\partial_1 S_{21}=\cos(\Omega(L - t_1)) -\Omega (t_2-L)  \sin(\Omega(L - t_1)) \, ,
\end{align}
leading to
\begin{equation}
  {\rm Cot}_{12}=\frac{C_{21}}{S_{12}}=-\frac{\Omega - \Omega^2(t_2-L)\tan \Omega (L-t_1)}{\Omega(t_2-L)+\tan \Omega (L-t_1)} \, .
\end{equation}
Combining all these results, the auxiliary functions now read
\begin{align}
\khat^2_{21} &= -2i\omega {\rm Cot}_{21}=\frac{-2i\omega \Omega}{\Omega (t_2-L) + \tan (\Omega(L-t_1))} \, ,\\
\jhat_{21} &= 2i\omega \Omega \cos^2(\Omega(L-t_1)) \big[\tan(\Omega(L-t_1))-\cot(\Omega t_1) \big] \, , \\
\rhat_{21} &= \frac{2i\omega }{ \khat^2_{21}S_{12}} =\frac{1}{\cos (\Omega(L-t_1))} \,.
\end{align}
Inserting these expressions into Eq.~\eqref{eq:IOE_NLO_INPOUT_1}, one realizes that the remaining $t_2$ integral can be carried out 
\begin{align}
\int_L^\infty \rmd t_2 \,\rme^{-\frac{k_\perp^2}{\khat^2_{21}}} =\int_L^\infty \rmd t_2 \, \rme^{-i\frac{k_\perp^2}{2\omega}\left[(t_2-L)+\frac{\tan(\Omega(L-t_1))}{\Omega}\right]}  =\frac{2\omega}{ik_\perp^2} \rme^{-i\frac{k_\perp^2}{2\omega\Omega}\tan(\Omega(L-t_1))} \, .
\end{align}
As a consequence, the \textbf{in-out} contribution to the NLO spectrum can be finally written as 
\begin{align}\label{eq:In_out_NLO_final}
(2\pi)^2\omega\frac{\rmd I_{\text{in-out}}^{\rm NLO}}{\rmd \omega \rmd^2 \k} &= \frac{2\bar{\alpha}\hat{q}_0\pi}{k_\perp^4} \Re \int_0^{L} \rmd t_1 \,  \cos^2(\Omega(L-t_1)) \nn
&\times I_b\left(\frac{k_\perp^2}{\jhat_{21}},\frac{k_\perp^2}{Q_r^2 \cos^2(\Omega(L-t_1))}\right) \rme^{-i \frac{k_\perp^2}{2\omega \Omega}\tan(\Omega(L-t_1))} \, .
\end{align}

\subsection{Final formulas}\label{sec:summary}

At this point, we summarize the main results obtained in the two previous sections. Our aim is to provide a set of compact equations which can be directly used in phenomenological studies or implemented in jet quenching Monte-Carlo codes. In what follows, we first present the results for a generic medium profile and then take the brick limit. 

\subsubsection{Spectrum at LO+NLO for a generic medium profile}

Up to NLO in the IOE, the purely medium-induced gluon spectrum, i.e. after subtracting vacuum radiation, can be written as
\begin{align}\label{eq:spec_res_generic_1}
\frac{\rmd I^{\rm LO+NLO}}{\rmd \omega  \rmd^2 \k} = \frac{\rmd I^{\rm LO}}{\rmd\omega \rmd^2 \k}+\frac{\rmd I_{\text{in}}^{\rm NLO}}{\rmd \omega \rmd^2 \k} +\frac{\rmd I_{\text{broad}}^{\rm NLO}}{\rmd \omega \rmd^2 \k}  \, .
\end{align}
The LO term reads
\begin{align}\label{eq:spec_res_generic_2}
    (2\pi)^2\omega\frac{\rmd I^{\rm LO}}{\rmd \omega \rmd^2 \k}&=8\Bar{\alpha} \pi \, \Re  \int_0^\infty \rmd t  \, {\rm Cot}(t) \frac{\rme^{-\frac{k_\perp^2}{\khat^2(t,0)} } }{\khat^2(t,0)} -\frac{8\pi \abar}{k_\perp^2} \, .
  \end{align}
 In addition, the NLO contributions are given by
\begin{align}\label{eq:spec_res_generic_3}
(2\pi)^2\omega\frac{\rmd I_{\text{in}}^{\rm NLO}}{\rmd \omega \rmd^2 \k} &=\frac{\bar{\alpha}\pi}{2\omega k_\perp^4} \Re\, i\int_0^\infty \rmd t_2 \,\int_0^{t_2} \rmd t_1 \,  \frac{\hat{q}_0(t_1)}{\rhat_{21}^2} \rme^{-\frac{k_\perp^2}{\khat^2_{21}}} 
\nn
&\times  \left[Q^2_s(\infty,t_2) I_a\left(\frac{k_\perp^2}{\jhat_{21}},\frac{k_\perp^2\rhat^2_{21}}{Q_r^2}\right)+2 C_{12}\rhat_{21} k_\perp^2 I_b\left(\frac{k_\perp^2}{\jhat_{21}},\frac{k_\perp^2\rhat^2_{21}}{Q_r^2}\right)\right] \, ,
\end{align}
and
\begin{align}\label{eq:spec_res_generic_4}
(2\pi)^2\omega\frac{
  \rmd I_{\rm \text{broad}}^{\rm NLO}}{\rmd \omega \rmd^2 \k}&= -\frac{\pi \bar{\alpha}}{k_\perp^4} \Re\int_0^\infty \rmd t_1 \, {\rm Cot}(t_1)\, Q^2_{s0}(\infty,t_1) \, I_a\left(\frac{k_\perp^2}{\khat^2(t_1)},\frac{k_\perp^2}{Q_b^2}\right)\, .
\end{align}
In the above equations we introduced 
\begin{align}
  {\rm Cot}(t_2,t_1)= {\rm Cot}_{21}\equiv \frac{C(t_1,t_2)}{S(t_2,t_1)}=\frac{C_{12}}{S_{21}} \,, 
\end{align}
where the $C$ and $S$ functions are described in \eqn{eq:Abel} and the functions $I_a$ and $I_b$ are given in Eqs.~\eqref{eq:bbS_alpha} and \eqref{eq:bbS_beta}, respectively. Further, the accumulated transverse momentum scales are $Q_{s0}^2(t_2,t_1)=\int_{t_1}^{t_2} \rmd t \, \hat{q}_0(t) $ and $Q_{s}^2(t_2,t_1) = Q_{s0}^2(t_2,t_1) \log\frac{Q_b^2}{\mu^2_\ast}$. The remaining functions are defined as follows,
\begin{align}
\khat^2_{21} &= Q^2_s(\infty,t_2) -2i\omega{\rm Cot}_{21} \, ,\nn
\jhat_{21} &=\frac{i}{2\omega}\left(-{\rm Cot}_{12}+{\rm Cot}_{10}+\frac{2i\omega}{S^2_{21}\khat^2_{21}}  \right)S_{21}^2 \khat_{21}^4\, , \nn 
\rhat_{21} & =-\frac{2i\omega}{\khat^2_{21}S_{21}} \, .
\end{align}
The matching scale $Q_b$, that enters everywhere into the ${\rm{\textbf{broad}}}$ and only into the $Q_s$ definition for the ${\rm{\textbf{in}}}$ case, is obtained by solving the transcendental equation
\beq\label{eq:Qb_final}
Q_b=\displaystyle\int_0^{\tilde L} \rmd t \,  \hat q_0(t)\log\frac{Q^2_b}{\mu^2_\ast} \, ,
\eeq
with $\tilde{ L}$ some effective medium length. The exact value of $\tilde{ L}$ is not important, as long as it is taken such that the relevant support for the integration of $\hat{q}$ is covered. The radiative matching scale, $Q_r$, appears in all other terms related to the kernel expansion and is the solution of
\beq\label{eq:Qr_final}
Q^2_r=\sqrt{\hat q_0(t)\omega \log\frac{Q^2_r}{\mu^2_\ast}} \, .
\eeq

\subsubsection{Spectrum at LO+NLO for the brick model}\label{subsubsec:LO_NLO_brick}

The previous results are simplified when the medium is modelled as a plasma brick of length $L$ with 
\begin{equation}
  \hat{q}(t)=\hat{q} \, \Theta(L-t) \, .
\end{equation}
In this case, the full medium-induced spectrum at NLO in the IOE can be written as 
\begin{align}\label{eq:spec_res_brick_1}
\frac{\rmd I^{\rm LO+NLO}}{\rmd \omega \rmd^2\k} = \frac{\rmd I_{\text{in-in}}^{\rm LO}}{\rmd \omega \rmd^2 \k}+ \frac{\rmd I_{\text{in-out}}^{\rm LO}}{\rmd \omega \rmd^2 \k}+\frac{\rmd I_{\text{in-in}}^{\rm NLO}}{\rmd \omega \rmd^2 \k}+\frac{\rmd I_{\text{in-out}}^{\rm NLO}}{\rmd \omega \rmd^2 \k} +\frac{ \rmd I_{\text{broad}}^{\rm NLO}}{\rmd \omega \rmd^2 \k}  \, .
\end{align}
The leading order terms read
\begin{align}\label{eq:spec_res_brick_2}
  (2\pi)^2\omega\frac{\rmd I^{\rm LO}_{\text{In-In}}}{\rmd \omega \rmd^2 \k}&=8\Bar{\alpha} \pi \, \Re \int_0^L \rmd  t  \, \Omega \, {\rm cot}(\Omega t)   \frac{\rme^{-\frac{k_\perp^2}{Q_s^2(L,t) -2i  \omega  \Omega {\rm cot}(\Omega t)} }}{Q_s^2(L,t) - 2i  \omega \Omega {\rm cot}(\Omega t)} \,, 
  \end{align}
and
\begin{align}\label{eq:spec_res_brick_3}
(2\pi)^2\omega\frac{\rmd I_{\text{In-Out}}^{\rm LO}}{\rmd \omega \rmd^2 \k}
&= \frac{8\bar{\alpha}\pi }{k_\perp^2}\Re  \left(\rme^{-\frac{ik_\perp^2}{2\omega \Omega \,  \cot\Omega L }   } -1\right)    \, .
    \end{align}
Here we used 
\begin{align}
 &\Omega  = \frac{1-i}{2}\sqrt{\frac{\hat{q}_0}{\omega} \log\frac{Q_r^2}{\mu^2_\ast}} \, , \\
&Q_s^2(L,t) = \hat{q}_0 \log\frac{Q_b^2}{\mu^2_\ast} (L-t_2)  \, .
\end{align}
 The NLO \textbf{in-in} term can be written as 
 \begin{align}\label{eq:spec_res_brick_4}
  (2\pi)^2\omega\frac{\rmd I_{\text{in-in}}^{\rm NLO}}{\rmd \omega \rmd^2 \k}&=\frac{\bar{\alpha}\pi\hat{q}_0}{2\omega k_\perp^4} \Re\, i\int_0^L \rmd t_2 \,\int_0^{t_2} \rmd t_1 \,  \frac{\rme^{-\frac{k_\perp^2}{\khat^2_{21}}}}{\rhat_{21}^2}  
  \nn
  &\times  \left[ Q^2_s(L,t_2) I_a\left(\frac{k_\perp^2}{\jhat_{21}},\frac{k_\perp^2\rhat^2_{21}}{Q_r^2}\right)+2 C_{12}\rhat_{21} k_\perp^2 I_b\left(\frac{k_\perp^2}{\jhat_{21}},\frac{k_\perp^2\rhat^2_{21}}{Q_r^2}\right)\right] \, ,
  \end{align}
where $Q^2_s(L,t_2)$ is defined as above, the $C$ and $S$ functions are given in Eq.~\eqref{eq:SC-brick}, and
\begin{align}
  & \khat^2_{21} = Q^2_s(L,t_2) -2i\omega \Omega \, {\rm cot}(\Omega (t_2-t_1))\, , \nn
  & \jhat_{21}=\frac{i }{2\omega}\left(\Omega{\rm cot}(\Omega(t_2-t_1))+\Omega{\rm cot}(\Omega t_1)+\frac{2i\omega}{S^2_{21}\khat^2_{21}}  \right)S_{21}^2 \khat_{21}^4\, , \nn 
  &\rhat_{21} =-\frac{2i\omega}{\khat^2_{21}S_{21}} \, .
 \end{align}
 The NLO \textbf{in-out} piece is
 \begin{align}
(2\pi)^2\omega\frac{\rmd I_{\text{in-out}}^{\rm NLO}}{\rmd \omega \rmd^2 \k} &= \frac{2\bar{\alpha}\hat{q}_0\pi}{k_\perp^4} \Re \int_0^{L} \rmd t_1 \,  \cos^2(\Omega(L-t_1)) \nn
&\times I_b\left(\frac{k_\perp^2}{\jhat_{21}},\frac{k_\perp^2}{Q_r^2 \cos^2(\Omega(L-t_1))}\right) \rme^{-i \frac{k_\perp^2}{2\omega \Omega}\tan(\Omega(L-t_1))} \, .
\end{align}
with
\begin{align}
 &\khat^2_{21}=\frac{-2i\omega \Omega}{\Omega (t_2-L) + \tan (\Omega(L-t_1))} \, ,\nn 
  &\jhat_{21}=2i\omega \Omega \cos^2(\Omega(L-t_1)) \big[\tan(\Omega(L-t_1))-\cot(\Omega t_1) \big]\, ,  \nn
&\rhat_{21} = \frac{1}{\cos (\Omega(L-t_1))} \, .
\end{align}
 Finally, the NLO \textbf{broad} contribution is given by
\begin{align}
  (2\pi)^2\omega\frac{
    \rmd  I_{\rm \text{broad}}^{\rm NLO}}{\rmd \omega \rmd^2 \k}
  &= -\frac{\hat{q}_0\pi \bar{\alpha}}{k_\perp^4} \Re \int_0^L \rmd t_2 \, \Omega \, {\rm cot}(\Omega t_2)\, (L-t_2) \,  I_a\left(\frac{k_\perp^2}{\khat^2(t_2,0)},\frac{k_\perp^2}{Q_b^2}\right) \, ,
  \end{align}
with 
\beq
 \khat^2(t_1,0)= \hat{q}_0 \log\frac{Q_s^2}{\mu^2_\ast}(L-t_1) -2i\omega\Omega{\rm cot}(\Omega t_1)\, .
\eeq

\subsection{Asymptotic behavior}\label{sec:asymp_beahvior}

The complete expressions for the in-medium branching kernel, that we have summarized in the previous section, are written in terms of a few integrations that we did not manage to solve analytically. Before presenting their numerical implementation, we would like to give further analytical insight into the discussion. To that end, we analyze the behavior of the IOE spectrum up to NLO in two physically relevant asymptotic regimes: when the emitted gluon is either (i) soft ($\omega\ll\omega_c$) and collinear ($k_\perp^2\ll \hat{q} L$) or (ii) hard ($\omega \gg \omega_c$) and wide angled ($k_\perp^2\gg \hat{q} L$). That is, the regime of validity of BDMPS-Z and GLV approaches, respectively. Our results below are obtained by taking the brick limit, although similar conclusions are obtained for other choices of medium profile. In addition, we neglect the purely vacuum radiation as it is completely irrelevant for this discussion. 

\subsubsection{Multiple soft scattering regime}\label{sec:MS_IOE}

We begin by analyzing the regime in which the emitted gluon is soft, i.e. $\omega \ll\omega_c$, and its typical formation time is much shorter than the medium length $t_f\sim \sqrt{\frac{\omega}{\hat q}}\ll L$. The latter condition can be translated into a constraint on the transverse momentum off the emission, $q_\perp^2\sim \hat{q} t_f \sim\sqrt{\hat{q}\omega }\ll \hat{q }L$,\footnote{Notice that this condition refers to the momentum of the in-medium vertex, rather than the final momentum of the gluon. Even for soft gluon emissions, final state broadening can lead to a final momentum $k_\perp^2\sim  \hat{q }L$. } which implies that short formation time gluons typically acquire most of their transverse momentum due to final state broadening. Under these conditions, the IOE spectrum simplifies significantly. The general formula for the medium-induced spectrum given by Eq.~\eqref{eq:spectrum} can be re-written in momentum space as\footnote{Strictly speaking, the upper bound of the integrals should be $L$. We take $L\to \infty$ to facilitate analytical manipulations.}
\begin{align}
\label{eq:didwdkt-ms}
  (2\pi)^2\omega\frac{\rmd I}{\rmd \omega \rmd^2 \k}&=\frac{2\bar{\alpha}\pi}{\omega^2} \Re\bigg[\int_0^\infty \rmd t_2 \int_0^{t_2} \rmd \tau \int_{\x,\q}  \rme^{-i \q\cdot \x}\cP(\k-\q;L-t_2)
  \nn &\times\bdel_\y\cdot \bdel_\x \, \cK(\x,t_2;\y,t_2-\tau)_{\y=0} \bigg] \, .
  \end{align}
  where we have exploited that $\cK$ and $\cP$ are invariant under time translations, i.e. depend only on time differences, when the plasma is homogenous. Eq.~\eqref{eq:didwdkt-ms} can be simplified noting that $\tau \sim t_f \ll t_2$ and thus one can set $t_2\to \infty$ in the $\tau$ integration upper limit. That is, the two time integrations decouple. In addition, we note that $q_\perp \sim1/x_\perp$ corresponds to the transverse momentum acquired in the branching process, $q_\perp \sim\sqrt{\hat{q}\omega}$, that, as we have anticipated, is small with respect to the characteristic broadening momentum, i.e. $q_\perp  \ll \hat{q}L$. As a consequence, we neglect $\q$ with respect to $\k$ inside $\cP$. Then, the $\q$ integral acts solely on $\cK$ and the $\x$ and $\q$ integrals yield
\begin{align}
\label{eq:factorized}
  (2\pi)^2\omega\frac{\rmd I}{\rmd \omega \rmd^2 \k}&=\frac{2\bar{\alpha}\pi}{\omega^2} \Re\bigg[\int_0^\infty \rmd t_2 \int_0^{\infty}  \rmd \tau \,  \cP(\k;L-t_2) \bdel_\y\cdot \bdel_\x \cK(\x,t_2;\y,t_2-\tau)_{\x=\y=0} \bigg] \, .
  \end{align}
A familiar element in the previous equation is the medium induced rate defined as
\beq
\omega\frac{\rmd I}{\rmd \omega \rmd t_2} &=& \frac{2\bar{\alpha}\pi}{\omega^2}\Re\bigg[ \int_0^{\infty} \rmd \tau \, 
\bdel_\y\cdot \bdel_\x \, \cK(\x,t_2;\y,t_2-\tau)_{\x=\y=0} \bigg] \, .
\eeq
Then, Eq.~\eqref{eq:factorized} can be finally written as
\beq\label{eq:ll_2}
 (2\pi)^2\omega\frac{\rmd I}{\rmd \omega \rmd^2 \k}= \int_0^\infty \rmd t_2 \,  \mathcal P(\k; L-t_2) \,  \omega\frac{\rmd I}{\rmd \omega \rmd t_2} \, .
\eeq
That is, in the soft and collinear limit, the spectrum is given by the product of the time integral of the broadening distribution and the medium induced energy rate. This result is not tied with the IOE approach and has been previously obtained in the literature in the context of BDMPS-Z calculations~\cite{BDIM1} and exploited in Monte-Carlo simulations~\cite{Saclay,Blanco:2020uzy}. We proceed to compute Eq.\eqref{eq:ll_2} in the IOE.

The soft limit of the IOE energy spectrum at all orders was computed in Refs.~\cite{IOE1,IOE3} and reads
\begin{align}\label{eq:ll_1}
\omega \frac{ \rmd I^{\rm IOE}}{\rmd \omega}=  \omega \frac{\rmd I^{\rm LO}}{d\omega}(\hat{q}\to \hat{q}_{\rm eff}) \, ,
\end{align}
where $\omega \frac{\rmd I^{\rm LO}}{\rmd \omega}= \bar{\alpha }\sqrt{\frac{\hat{q}L^2}{\omega}}$ corresponds to the well known BDMPS-Z result. The effective jet quenching parameter is given at leading-logarithmic order by~\cite{IOE3}\footnote{As shown in Ref.~\cite{IOE3}, if all terms in $\hat{q}_{\rm eff}$ are resummed, Eq.~\eqref{eq:ll_1} gives the full energy spectrum for $\omega \ll \omega_c$.}
\begin{align}\label{eq:ll_3}
 \hat{q}_{\rm eff}=\hat{q}_0 \log\left(\frac{Q_r^2}{\mu_\ast^2}\right) \, \left(1+\frac{1.016}{\log\left(\frac{Q_r^2}{\mu_\ast^2}\right)}+\mathcal{O}\left(\frac{1}{\log^2\left(\frac{Q_r^2}{\mu_\ast^2}\right)}\right)\right) \, .
\end{align}
Eqs.~\eqref{eq:ll_1} and \eqref{eq:ll_3} show that the IOE energy spectrum is governed by the LO result with higher orders suppressed by a logarithmic power which can be written in terms of the ratio $\frac{\hat{q}_0}{\hat{q}}$. A similar conclusion can be reached regarding the broadening distribution in the kinematical limit $k_\perp^2\ll \hat{q}L$, see for example \eqn{eq:assist_1} for the result up to NLO. Combining these two results, one concludes that the soft and collinear limit of the fully differential spectrum in Eq.~\eqref{eq:ll_2} also obeys this functional form. Note that the spectrum will consist of terms where the matching scale is given by $Q_r$ and others where it is $Q_b$, depending if the terms come from the expansion of the kernel or of the broadening distribution.

In Appendix~\ref{app:In_Out_example} we explicitly show that the \textbf{in-out} NLO contribution in the IOE spectrum scales as the LO term multiplied by a logarithm that arises from the ratio $\frac{\hat{q}_0}{\hat{q}}$.

\subsubsection{Rare, hard scattering regime}

Let us now consider the orthogonal regime with respect to the previous section. Here, the gluon is hard, $\omega\gg \omega_c$, and carries a large transverse momentum $k_\perp^2\gg \hat{q} L$, i.e. the intrinsic momentum of the gluon is significantly larger than what it typically acquires through broadening in the medium. In this case, the multiple soft scattering contribution is suppressed by the LPM effect and the emission spectrum is dominated by single hard scattering in the medium. This corresponds to the truncation of the opacity expansion, considered by GLV, at first order~\cite{GLV,Wiedemann}, leading to an emission spectrum reading
\begin{equation}\label{eq:glv}
\begin{split}
(2\pi)^2\omega\frac{\rmd I^{\rm GLV}}{\rmd \omega \rmd^2 \k} &= \frac{2\Bar{\alpha}\hat{q}_0L^3\pi}{\omega^2}\int_0^\infty \rmd x \, \frac{x-\sin(x)}{x^2} 
\frac{\gamma+u-x}{(u^2+2u(\gamma-x)+(\gamma+x)^2)^{3/2}} \, ,
\end{split}
\end{equation}
where $u=\frac{L}{2\omega}k_\perp^2$ and $\gamma=\frac{\mu^2L}{2\omega}$. The medium potential was taken to be the GW model and thus $\mu$ is its infrared regulator. It can be related to the universal physical mass $\mu_\ast$, as mentioned in Section~\ref{sec:momentum-broadening} and detailed in Refs.~\cite{broadening_paper,IOE3}. 

Note that the conditions $\omega \gg \omega_c$ and $k_\perp^2\gg \hat q L$ correspond to $u\gg 1 \gg \gamma$ in the previous equation. To take this limit in Eq.~\eqref{eq:glv}, let us consider the integral
\begin{equation}\label{eq:glv_app1}
\begin{split}
I &\equiv \int_0^\infty \rmd x \, \frac{x-\sin(x)}{x^2} 
\frac{\gamma+u-x}{(u^2+2u(\gamma-x)+(\gamma+x)^2)^{3/2}} \, ,
\end{split}
\end{equation}
in two regions: (i) $u\gg x\gg \gamma$ ($<$) and (ii) $u\sim x\gg 1$, but $u-x\gg \gamma$ ($>$). First, we split $I$ using
\begin{equation}
\int_0^\infty \to \lim_{\epsilon \to 0} \, \int_0^{\epsilon u} + \int_{\epsilon u}^\infty \equiv I_< + I_> \,,   
\end{equation}
with $\epsilon u$ held constant. The contribution from the first region can be easily computed to leading-logarithmic accuracy
\begin{equation}
\begin{split}
I_<&= \lim_{\epsilon \to 0} \, \int_0^{\epsilon u} \rmd x \,  \frac{x-\sin(x)}{x^2} 
\frac{1}{u^{2}} + \mathcal{O}\left(\frac{x}{u}\right) =\lim_{\epsilon \to 0} \, \frac{1}{u^{2}} G(\epsilon u)\approx \frac{1}{u^2}\left[\log(\epsilon u)-1+\gamma_E\right]  \, ,
\end{split}    
\end{equation}
where we introduced\footnote{${\rm Ci}(x)=-\int_x^\infty \rmd t \, \frac{\cos (t)}{t}$.}
\begin{align}\label{eq:G}
G(a)\equiv \int_0^{a} \rmd x \, \frac{x-\sin(x)}{x^2}=\log(a)-1+\gamma_E- {\rm Ci}(a)+\frac{\sin(a)}{a} \, .  
\end{align}
In the case of $I_>$, we first notice that $x\gg1$ so we can drop the $\sin(x)$ term. Defining $a\equiv \gamma/u\ll 1$ we have
\begin{equation}
\begin{split}
I_>&\approx \frac{1}{u^2}\int_{\epsilon u}^\infty \frac{\rmd z}{z} \frac{1-z}{((1-z)^2+4za)^{\frac{3}{2}}}  \\
&=\frac{1}{u^2}\left[\frac{2(4a-3+z)}{4(a-1)\sqrt{z^2+1+(4a-2)z}}-\arctan\left(\frac{1+(2a-1)z}{\sqrt{z^2+(4a-2)z+1}}\right)\right]_{\epsilon u}^\infty\\
&=\frac{1}{u^2}\left[-2+\frac{1}{2}\log \left(\frac{1-a}{a}\frac{-1}{a(a-1)(\epsilon u)^2}\right)\right]=\frac{1}{u^2}\left(-2+\log\left(\frac{u}{\gamma}\right)-\log(\epsilon u)\right)
\, .
\end{split}
\end{equation}
Combining the results from the two different regions, the $I$ integral gives
\begin{equation}
I=\frac{1}{u^2}\left(-3+\gamma_E+\log\left(\frac{u^2}{\gamma}\right)\right)+ \mathcal{O}\left(\frac{1}{u^3}\right)    \, .
\end{equation}
Inserting this result into the GLV spectrum given in Eq.~\eqref{eq:glv} yields
\begin{equation}\label{eq:glv_largekt}
\begin{split}
(2\pi)^2\omega\frac{\rmd I^{\rm GLV}}{\rmd \omega \rmd^2 \k} &\approx \frac{2\bar{\alpha}\hat{q}_0L^3\pi}{\omega^2}\frac{1}{u^2}\left(\log\left(\frac{u^2}{\gamma}\right)+\gamma_E-3 \right)=\frac{8\bar{\alpha}\hat{q}_0L\pi}{k_\perp^4}\log\left(\frac{k_\perp^4L \rme^{\gamma_E-3}}{2\omega \mu^2}\right)\, .
\end{split}
\end{equation}
The expected $1/k_\perp^4$ power tail naturally arises from a Coulomb-like single hard scattering in the medium. Counterintuitively, even though we are considering here the high energy limit, the spectrum is still sensitive to the infrared details of the in-medium scattering potential via the thermal mass $\mu$. For the sake of comparing with the IOE in what follows, a couple of manipulations are required. First, we re-write the resulting logarithm as
\begin{equation}
  \log\left(\frac{k_\perp^4L \rme^{\gamma_E-3}}{2\omega \mu^2}\right)=\left(\gamma_E-3+\log\left(\frac{k_\perp^2}{\mu^2}\right)+\log\left(\frac{k_\perp^2L}{2\omega}\right)\right)\, ,
  \end{equation}
and then replace the GW mass by the universal infrared scale $\mu_\ast$ through the leading-logarithmic prescription $4\mu_\ast^2=\mu^2 \rme^{-1+2\gamma_E}$~\cite{broadening_paper,IOE3}. This leads to our final expression for the GLV spectrum:
\begin{equation}
    \begin{split}
    (2\pi)^2\omega\frac{\rmd I^{\rm GLV}} {\rmd \omega \rmd^2 \k}=\frac{8\bar{\alpha}\pi\hat{q}_0L}{ k_\perp^4}\left(3\gamma_E-4+\log\left(\frac{k_\perp^2}{4\mu_\ast^2}\right)+\log\left(\frac{k_\perp^2L}{2\omega}\right)\right)\, . 
    \end{split}
    \end{equation}

Let us now take the $\omega\gg \omega_c$ and $k_\perp^2\gg \hat{q}L$ limits in the IOE spectrum. At leading order, since $k_\perp^2\gg \hat{q}L \sim Q_s^2$, broadening contributions are sub-leading and thus can be ignored in Eq.~\eqref{eq:HO_brick_InIn}. Further, $\omega \gg \omega_c$ is equivalent to $\Omega L \ll 1$. Combining this pair of observations yields the LO contributions at $\mathcal O(\k^2)$:
\begin{align}\label{eq:oo_1}
(2\pi)^2\omega\frac{\rmd I^{\rm LO}_{\rm \text{in-in}}}{\rmd \omega \rmd^2 \k}&\approx\frac{4\bar{\alpha} \pi}{\omega} \, \Re \bigg[ \, \int_0^L \rmd t  \,  \exp\left[-\frac{ik_\perp^2}{ 2 \omega } t\right] \bigg]=\frac{8\Bar{\alpha}\pi}{k_\perp^2}\left(1-\cos \frac{k_\perp^2L}{2\omega }\right)\,, 
\end{align}
and
 \begin{align}
(2\pi)^2\omega\frac{\rmd  I_{\rm \text{in-out}}^{\rm LO}}{\rmd \omega \rmd^2 \k}
&\approx -\frac{8\Bar{\alpha}\pi}{k_\perp^2}\left(1-\cos \frac{k_\perp^2L}{2\omega }\right)\,. 
\end{align}
Adding the two components results into a vanishing spectrum at this order in $\k$ and indicates the need to go to higher orders that, as we will see, affect rather differently the \textbf{in-in} and \textbf{in-out} terms. The latter is exponentially suppressed when including higher orders as can be derived from Eq.~\eqref{eq:HO_brick_out_2}. In turn, the \textbf{in-in} term follows a power-law suppression. To see this, we perform a second order gradient expansion of the broadening distribution in the $Q_s^2\ll k_\perp^2$ limit as
\beq\label{eq:grad_exp}
\int_\p  \cP^{\rm LO}(\k-\p) u(\p) &=& \int_\q  \cP^{\rm LO}(\q) u(\k-\q)  \nn
&\approx& \int_\q \cP^{\rm LO}(\q) \left[ 1+ \q^i \nabla^i _\k  +\frac{1}{2} \q^i\q^j  \nabla^i _\k \nabla^j _\k \right] u(\k)\nn
&=&  \Bigg[ 1+ \frac{1}{4} \underbrace{\int_\q  q_\perp^2 \cP^{\rm LO}(\q) }_{\hat{q}L}  \nabla^2 _\k  \Bigg] u(\k) \, ,
\eeq
where $u(\p)$ is a test function, we have used unitarity in the first line and rotational symmetry to drop the linear term. The first term in brackets corresponds to the result already obtained in Eq.~\eqref{eq:oo_1}, where the broadening distribution was replaced by a Dirac $\delta$-function. Keeping only the second term in Eq.~\eqref{eq:grad_exp} and plugging it into Eq.~\eqref{eq:HO_brick_InIn} we obtain

\begin{align}
(2\pi)^2\omega\frac{\rmd I_{\rm \text{in-in}}^{\rm LO}}{\rmd \omega \rmd^2 \k}&\approx\frac{\bar{\alpha}\pi\hat{q}L}{\omega}  \nabla^2_\k \Re \bigg[i \, \int_0^L  \rmd t_2 \, 
  \rme^{-\frac{i \k^2 \tan(\Omega t_2) }{2 \omega \Omega }  }\bigg]   \, .
\end{align}
Now, because $\k$ is large the phase oscillates rapidly unless $t_2$ is small enough. To estimate the support of the $t_2$ integral, we exploit the fact that in the high energy regime $\Omega L \ll 1$ and thus the dominant contribution to the integral comes from the region where $t_2\ll \frac{2\omega}{k_\perp^2}\ll \Omega^{-1}$. As a consequence, one can replace the integration limit $L\to \infty$ and linearize the tangent, to obtain the leading asymptotic behavior of the spectrum
\begin{align}
\label{eq:lo-highenergy}
  (2\pi)^2\omega\frac{\rmd I_{\rm \text{in-in}}^{\rm LO}}{\rmd \omega \rmd^2 \k}&\approx\frac{\bar{\alpha}\pi\hat{q}L}{\omega}   \partial_\k^2\left(4k_\perp^2 \partial_\k^2\right)\Re \bigg[i \, \int_0^\infty \rmd t_2 \, 
    \rme^{-i \frac{\k^2 t_2}{2\omega } } \bigg] = \frac{8\bar{\alpha} \pi \hat{q}_0L}{k_\perp^4}\log \frac{Q_b^2}{\mu_\ast^2} \, .
  \end{align}
Notice that the logarithm depends on the broadening matching scale since it originates from the last line in Eq.~\eqref{eq:grad_exp}. Interestingly, the leading order contribution exhibits a $1/k_\perp^4$ tail, physically corresponding to early time hard emissions, which then suffer multiple scatterings in the medium, acquiring a momentum $\hat{q}L$ much smaller than the momentum off the emission vertex. Compared to the vacuum like emission in Eq.~\eqref{eq:oo_1}, we observe that although final state broadening does not change the power-law dependence on the transverse momentum, it power suppresses this second order by a $\frac{\hat{q}L}{k_\perp^2}\ll 1$ factor.

At NLO, we need to analyze individually each of the three identified terms in this asymptotic regime. Starting with the \textbf{broad} term, we note that in the high energy limit 
\beq
  \khat^2(t,0) \simeq  \hat q (L-t) -\frac{2i\omega }{t}\simeq -\frac{2i\omega }{t} \, ,
\eeq
so that
\begin{align}\label{eq:IOE_med_broad_app_3}
&(2\pi)^2\omega\frac{
  \rmd  I_{\rm \text{broad}}^{\rm NLO}}{\rmd \omega \rmd^2 \k}= -\frac{\pi \bar{\alpha}\hat q_0}{k_\perp^4} \Re\bigg[\,\int_0^L \frac{\rmd  t }{t}\, \, \, (L-t) \, I_a\left(i\frac{k_\perp^2}{ 2\omega}t,\frac{k_\perp^2}{Q_b^2}\right)\bigg]\,.
\end{align}
It is convenient to write the remaining integral as 
\beq
&& \int_0^L \frac{\rmd t }{t}\, \, \, (L-t) \,  I_a\left(i\frac{k_\perp^2}{ 2\omega}t,\frac{k_\perp^2}{Q_b^2}\right)=  \frac{2\omega}{ik_\perp^2}\int_0^{x_{\rm max}} \frac{\rmd x }{x}\, \, \, (x_{\rm max}-x) \, I_a\left(x,y \right)\, ,
\eeq
where we used $x = i\frac{k_\perp^2}{2\omega}t$, $y= \frac{k_\perp^2}{Q_b^2}$ and $x_{\rm max}= i\frac{k_\perp^2}{2\omega}L\gg1$. This simplified integral is analytically solvable
\begin{align}
&\int_0^{x_{\rm max}}  \frac{\rmd x}{x}(x_{\rm max}-x)I_a(x,y)=-8\Bigg(-x^2+2\rme^{-x}(x_{\rm max}-6-2x)+(x_{\rm max}-4)\left[\log(x)\right.  \nn &\left.+x-2{\rm Ei}(-x)\right]+\rme^{-x}\left[4(1+x)+2x^2+x^3-x_{\rm max}(1+x+x^2)\right]\left[{\rm Ei}(x)-\log\frac{4x^2}{y}\right] \Bigg)    \, .
\end{align}
Truncating the previous exact result to leading order in $x_{\rm max}$, we obtain
\beq
&&\int_0^{x_{\rm max}} \frac{\rmd x }{x}\, \, \, (x_{\rm max}-x) \, I_a\left(x,y \right) \approx 8 \left[  \log \frac{4}{yx_{\rm max}}  + 5 - 3 \gamma_E    \right] x_{\rm max}    \, .
\eeq
Plugging this last result into Eq.~\eqref{eq:IOE_med_broad_app_3} yields  
\begin{align}\label{eq:IOE_med_broad_app_4}
&(2\pi)^2\omega\frac{
  \rmd I_{\rm \text{Broad}}^{\rm NLO}}{\rmd \omega \rmd^2 \k}\approx  \frac{8\pi \bar{\alpha}\hat q_0 L}{k_\perp^4}  \left[  \log \frac{k_\perp^2}{4Q_b^2}+ \log \frac{k_\perp^2}{2\omega}L - 5 +3 \gamma_E  \right] \,.
\end{align}
We restrain ourselves from doing any physical interpretation of this result at this stage and proceed to compute the \textbf{in-in} contribution. In this case, the lack of a vacuum-like ($\rmd t/t$) divergence simplifies the calculation. First, we use the asymptotic form of the exponential integral function,
\begin{align}
  {\rm Ei}(x)=  \frac{\rme^x}{x}\sum_{n=0}^{N-1} \frac{n!}{x^n}+\mathcal{O}\left(\frac{N!}{x^N}\right)  \, ,
\end{align}
 in Eqs.~\eqref{eq:bbS_alpha} and \eqref{eq:bbS_beta}, to obtain the leading asymptotic forms of $I_a$ and $I_b$
\beq
I_a \approx -8 \,, \quad  I_b \approx 4 \, .
\eeq 
Moreover, since $\Omega L\ll 1$ we can simplify the auxiliary functions given by Eq.~\eqref{eq:help_2} down to
\beq
 \rhat_{21} = C_{12}\approx 1 \, , \quad \khat_{21}^2\approx -\frac{2i\omega}{(t_2-t_1)}\,.
\eeq
Consequently, the \textbf{in-in} spectrum reduces to
\begin{align}
(2\pi)^2\omega\frac{\rmd I^{\rm NLO}_{\rm \text{in-in}}}{\rmd \omega \rmd^2 \k}&\approx \frac{\pi \bar{\alpha}\hat q_0}{2\omega k_\perp^4} \Re\bigg[ i\int_0^L \rmd t_2 \,\int_0^{t_2} \rmd t_1 \,  \rme^{-i\frac{ k_\perp^2}{2\omega}(t_2-t_1)}   \left(-8\hat q (L-t_2) +8  k_\perp^2\right)  \bigg] \nn 
&\approx \frac{4\pi \bar{\alpha}\hat q_0}{\omega k_\perp^2} \Re\bigg[ i\int_0^L \rmd t_2 \,\int_0^{t_2} \rmd t_1 \,  \rme^{-i\frac{k_\perp^2}{2\omega}(t_2-t_1)} \bigg]\, \nn
& = \frac{8\pi \bar{\alpha}\hat q_0L}{k_\perp^4} +\cO(k_\perp^{-6})\,.
\end{align}
Finally, for the \textbf{in-out} term we obtain
\begin{align}
&(2\pi)^2\omega\frac{\rmd I_{\rm \text{in-out}}^{\rm NLO}}{\rmd \omega \rmd^2 \k}\approx\frac{8\bar{\alpha}\hat{q}_0\pi}{k_\perp^4} \Re\bigg[ \int_0^{L} \rmd t_1 \,   \rme^{-i \frac{k_\perp^2}{2\omega }t_1}  \bigg]=\frac{16\bar{\alpha}\hat{q}_0\pi \omega}{k_\perp^6}\sin\left(\frac{k_\perp^2L}{2\omega}\right) \, ,
\end{align}
which is power-suppressed with respect to the \textbf{broad} and \textbf{in-in} terms and can be ignored. Then, the NLO contribution to the IOE spectrum is given by
\begin{align}\label{eq:IOE_large_kt}
&(2\pi)^2\omega\frac{\rmd I^{\rm NLO}}{\rmd \omega \rmd^2 \k}\approx\frac{8\pi \bar{\alpha}\hat q_0L}{k_\perp^4}\left[3 \gamma_E-4+\log\left(\frac{k_\perp^2}{4Q_b^2}\right)+\log\left(\frac{k_\perp^2L}{2\omega}\right)\right]\,.
\end{align}
Finally, combining the LO and NLO results we obtain that the IOE spectrum at high energies reduces to 
\begin{align}
\label{eq:final-highenergy}
  (2\pi)^2\omega\frac{\rmd I^{\rm LO+NLO}}{\rmd \omega \rmd^2 \k}&=\frac{8\pi \bar{\alpha}\hat q_0L}{k_\perp^4}\left[3 \gamma_E-4+\log\left(\frac{k_\perp^2}{4Q_b^2}\right)+\log\left(\frac{k_\perp^2L}{2\omega}\right)+\log\left( \frac{Q_b^2}{\mu_\ast^2} \right)\right]
  \nn &= \frac{8\pi\bar{\alpha}\hat{q}_0L}{ k_\perp^4}\left[3\gamma_E-4+\log\left(\frac{k_\perp^2}{4\mu_\ast^2}\right)+\log\left(\frac{k_\perp^2L}{2\omega}\right)\right]\nn &= (2\pi)^2\omega\frac{\rmd I^{\rm GLV}} {\rmd \omega  \rmd^2 \k}\,.
  \end{align}
 A few important remarks are in order at this point. The most obvious one is that the LO+NLO exactly matches the GLV result in the large $\omega$, large $\k$ regime. This was somehow expected but not trivial to confirm explicitly. Among other technicalities, a second order gradient expansion in transverse momentum for the LO term was essential. Another remarkable and related feature of the final result is that both the LO and NLO depend on $Q_b$, while their sum does not. This was already encountered in the energy spectrum calculation as discussed in Ref.~\cite{IOE3} and constitutes a sanity check of the Improved Opacity Expansion framework. We note again that unlike the soft limit considered before, this regime, although having a non-trivial cancellation of the matching scale dependence between different orders, does not provide any constraint on the functional form of $Q_b$, as can be observed from Eq.~\eqref{eq:final-highenergy}. This is unlike, for example, the result in Eq.~\eqref{eq:IOE_smallkt_final}, that forbids the matching from being a numerical constant. From a more pragmatic point of view, the exact matching between the IOE and GLV provides a non-trivial check on the computations performed in the previous sections.

\section{Numerical results}\label{sec:numerics}

In this section, we numerically explore the IOE spectrum in the brick model. We compare our results of the IOE spectrum truncated at LO and LO+NLO (see Section~\ref{subsubsec:LO_NLO_brick}) to (i) the single, hard scattering limit encompassed in the GLV spectrum (see Eq.~\eqref{eq:glv}) and (ii) an all-orders resummation of the spectrum presented in~\cite{CarlotaFabioLiliana}. Notice that the LO result can be considered as the BDMPS-Z solution with ultraviolet regulator taken to be the radiative matching scale $Q_r$. 

These comparisons should be regarded, at this point, as a merely theoretical exercise. However, we choose the medium parameters to be in the ballpark of LHC conditions. More concretely, the LHC-inspired medium has: $\hat q_{0}\!=\!0.156$~GeV$^3$, length $L\!=\!6$~fm and infrared regulator $\mu_\ast\!=\!0.355$~GeV. Further, we take a fixed value of the strong coupling constant $\alpha_s\!=\!0.28$ and consider radiation off a hard quark such that $C_R=C_F\!=\!4/3$. These set of parameters lead to a critical frequency scale $\omega_{c0}\!\equiv\!\hat q_0 L^2\!=\!140$~GeV and the saturation density $Q^2_{s0}\equiv \hat q_0 L\!=\!4.68$~GeV$^2$. Regarding our numerical routines, they run in a regular laptop with an average computing time of $\mathcal{O}(1)$ seconds for each pair of ($\omega,\k$) values, considering the above set of parameters. The computing time is significantly smaller, $\mathcal{O}(10^{-2})$ seconds, if not too extreme values of the kinematic variables are chosen.

\begin{figure}[ht]
  \centering
  \includegraphics[scale=0.45,page=2]{plot-Qs-variations.pdf}   \includegraphics[scale=0.45,page=1]{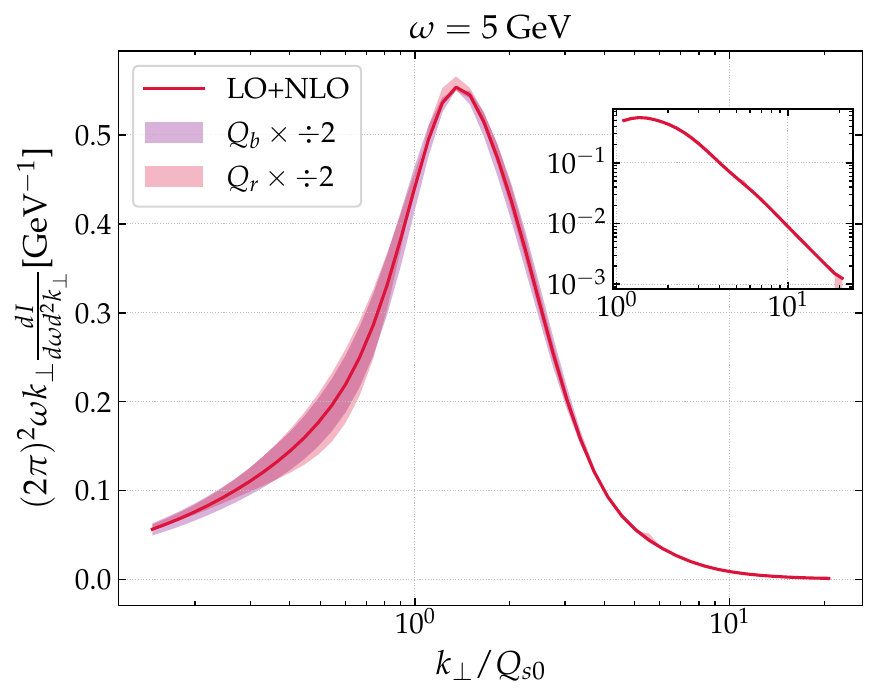} \\
    \includegraphics[scale=0.45,page=4]{plot-Qs-variations.pdf}   \includegraphics[scale=0.45,page=3]{plot-Qs-variations.pdf}
  \caption{Impact of variations by a factor of 2 in the two matching scales, $Q_b$ and $Q_r$, on the LO (left column) and the LO+NLO (right column) at two different frequencies: $\omega\!=\!5, 100$ GeV on the top and bottom rows, respectively. }
  \label{fig:qs-variations}
\end{figure}

Before comparing our result to other approaches, we first address a natural question: what is the dependence of the IOE spectrum on the matching scales $Q_r$ and $Q_b$? In Fig.~\ref{fig:qs-variations} we plot the medium-induced spectrum as a function of $k_\perp/Q_{s0}$ for two different gluon frequencies, $\omega\!=\!5, 100$ GeV, truncating the spectrum at LO (left) or at LO+NLO (right). The central curves are obtained by solving Eqs.~\eqref{eq:old_Qb} and \eqref{eq:Q_r}. Then, we perform an independent variation of the matching scales by a factor of $2$ ($1/2$) that lead to the uncertainty bands around the mid value. We recall that this variation is associated to the uncertainty in the definition of such scales, which are constrained up to an overall constant factor. 

Let us discuss first the large $\omega$, large $k_\perp$ regime, i.e. the inset in the bottom row plots. Analytically, we have shown that, in the asymptotic kinematical region, the dependence on the matching scale vanishes when one considers LO+NLO contribution, see Eq.~\eqref{eq:final-highenergy}. This is exactly what we observe numerically. Although this conclusion was reached for highly energetic gluons, the numerical results indicate that it holds reasonably well in the case of soft gluons too. Notice that when analyzing only the LO, a bigger, but still weak dependence on the matching scale $Q_b$ is observed. This corresponds to Eq.~\eqref{eq:lo-highenergy}, where a logarithmic dependence on $Q_b$ appears. Then, we reemphasize that only when considering LO+NLO the dependence on the matching scale is residual as due to the cancellation occurring between these two terms. 

The small $\omega$, small $\k$ scenario is represented by the top row plots in Fig.~\ref{fig:qs-variations}. In this region, we argued in the previous section that all orders scale as the LO term, with logarithmic power corrections as one goes higher in the IOE, see Eqs.~\eqref{eq:ll_2} and \eqref{eq:ll_1}. Numerically, we observe that there is a large uncertainty due to the variation of the matching scales at LO, mainly from $Q_r$. However, if one includes the NLO term (top right) the dependence on the matching scales almost disappears. Regarding $Q_b$, we note that higher orders in the broadening factor $\cP$ also enter through logarithmic power corrections on top of the LO term. Thus, the weaker dependence of LO+NLO on $Q_b$ as compared to LO is to be expected. These findings are inline with~\cite{IOE3}, where it was observed that for the energy spectrum the variations of the matching scale $Q_r$ although could drastically change the LO and the NLO terms separately, the sum LO+NLO was only sensitive to these variations at NNLO. This is a consequence of the fact once all orders in Eq.~\eqref{eq:ll_3} are considered, the spectrum becomes independent of $Q_r$. Since in Fig.~\ref{fig:qs-variations} the largest uncertainty comes from the scale $Q_r$, we argue that the result obtained here is a manifestation of the findings of~\cite{IOE3}. 

\begin{figure}[h!]
    \centering
    \includegraphics[scale=.78]{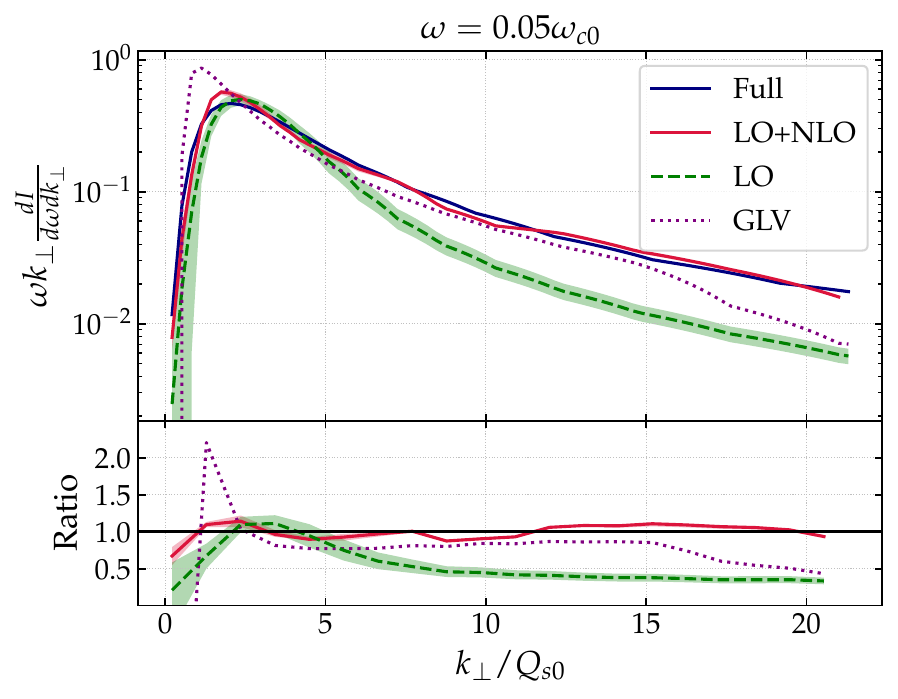}
    \includegraphics[scale=.78]{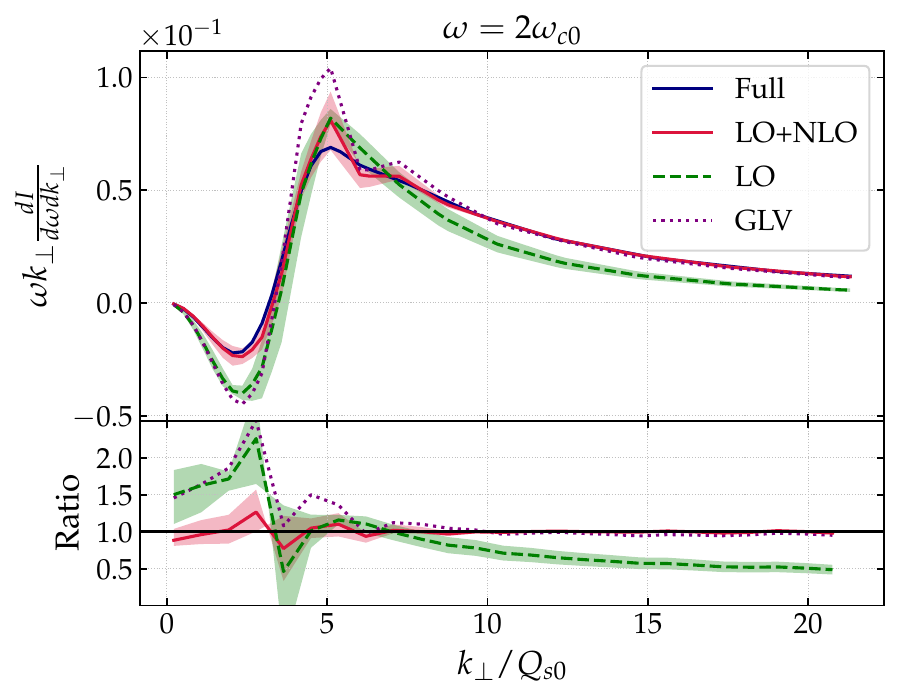}
    \caption{Comparison between the GLV spectrum (dotted, purple), the LO result (dashed, green), the IOE at LO+NLO (solid, red) and the all-order spectrum (solid, navy) as computed in Ref.~\cite{CarlotaFabioLiliana} for two gluon frequencies: $\omega= 0.05\, \omega_{c0}$ (left) and $\omega= 2\, \omega_{c0}$ (right). The ratio to the full solution is presented in the bottom panels. The uncertainty band arises from variations in the matching scale.}
    \label{fig:lhc}
\end{figure}

In Fig.~\ref{fig:lhc} we present the final comparison between the IOE spectrum and the other approaches mentioned above. We considered two gluon frequencies, $\omega= 0.05\, \omega_{c0}$ and $\omega= 2\, \omega_{c0}$, corresponding to the cases of soft and hard gluon and we use the GW mass $\mu^2\!=\!0.43$~GeV$^2$. Again, we vary the matching scales of the IOE spectrum up and down by a factor of two, independently and then take the envelope to build the uncertainty band.

The overall conclusion from the two plots in Fig.~\ref{fig:lhc}, and the most important result of this paper, is that the IOE spectrum, up to NLO, already does a reasonable job at capturing the full solution (less than 25\% deviations), with the advantage that it requires considerably less computational power. The observed deviation from the full numerical result reflects the sensitivity of the transverse momentum distribution to the infrared. A wider separation between $\mu^2$ and $Q_b^2$ or $Q_r^2$ would yield a better agreement. Let us split the discussion of Fig.~\ref{fig:lhc} into small and large transverse momentum.  

The small $k_\perp$ and small $\omega$ regime is dominated by multiple scattering contributions. Then, it is natural that the GLV spectrum fails to capture the full solution. The LO term of the IOE (related to the BDMPS-Z solution) already does a good job at describing the full result. Nonetheless, including the NLO term, improves not only the overall agreement with the full result, but also reduces the uncertainty band associated with the variation of the matching scales, as discussed in the previous section. When increasing the gluon's frequency, and still at small $k_\perp$, we observe that the LO+NLO result remarkably captures the full solution up to $5\%$ deviations. Regarding GLV, its agreement with the full solution is improved with respect to the small frequency case and, curiously, coincides with the LO term. We do not expect this to be a systematic result for other choices of the medium parameters. 

The large $k_\perp$ tail is generated by rare, hard scatterings in the medium. It is well known that the BDMPS-Z approximation does not correctly capture such contributions and, therefore, fails to describe the full solution. At small frequencies, we observe that the LO+NLO result approaches much faster the full solution than compared to GLV. This is to be expected, as at large, but not infinite $k_\perp$, multiple scatterings still play a role, despite being sub-leading. GLV lacks those effects and then needs an asymptotically large value of $k_\perp$ to reproduce the full solution, while the LO+NLO works even far from the asymptotic regime. At large frequencies, the spectrum is really dominated by a single hard scattering in the medium and thus the LO+NLO, full and GLV results rapidly converge. 

\section{Conclusion and Outlook}\label{sec:conclusion}

This work constitutes a natural extension of the recent studies of the medium-induced energy spectrum and broadening distribution using the Improved Opacity Expansion~\cite{IOE1,IOE2,IOE3,broadening_paper}. We have computed the in-medium radiative kernel using the IOE up to next-to-leading order accuracy, in the soft gluon approximation for a generic medium profile as well as for a brick plasma for which we have performed numerical 
computations that we compare to full numerical results from \cite{CarlotaFabioLiliana}. We observe a very good agreement for a LHC-motivated choice of medium parameters.     

From a theoretical viewpoint, the differential spectrum calculation highlights the role played by the matching scales that enters the definition of $\hat q$, given that it convolutes contributions due to final state broadening with in-medium radiative terms. Each of these physical processes enters the IOE expansion with its own matching scale that we denote by $Q_b$, associated to final state broadening terms, and $Q_r$, related to the radiative kernel. These two scales are obtained by solving their corresponding transcendental equations that are given, in full generality, by Eqs.~\eqref{eq:Qb_final} and \eqref{eq:Qr_final}. We emphasize that these two scales have to be treated separately in order for the expansion to be consistent and well defined. Taking this into account, we derive the medium-induced spectrum for a smooth medium time profile and also in the brick limit. The final formulas are given in Sec.~\ref{sec:summary}. 

Besides the master formulas, we analytically study the IOE for the plasma brick model in two asymptotic kinematical regimes. Firstly, we consider the regime where multiple soft exchanges with the medium constitute the dominant contribution, i.e. $\omega \ll \omega_c$, and with the further assumption that the gluon is collinear, i.e. $k_\perp^2\ll\hat{q}L$. In this case, we recover the well-known factorization formula given by Eq.~\eqref{eq:ll_2}, often used in jet quenching phenomenology~\cite{Saclay,BDIM1,Blanco:2020uzy,Kutak1}. In particular, this result implies that in the soft limit the differential spectrum can be written as the product between the LO term and powers of $\frac{\hat{q}_0}{\hat{q}}$ that correspond to higher order contributions. This result agrees with what was observed in the energy spectrum calculation, as detailed in~\cite{IOE3}. Secondly, we study the physically opposite regime where a single hard scattering with the medium governs the dynamics of the medium-induced spectrum. This corresponds to a region of phase space where $\omega \gg \omega_c$ and final momentum of the gluon satisfies $k_\perp^2\gg\hat{q}L$. In this regime, it is expected that the exact spectrum is given by the GLV result, and thus should be reproduced by the IOE approach. Indeed, we confirm that after considering both the LO and NLO contributions, non-trivial cancellations between these two orders occur such that one recovers the GLV spectrum. Not only the GLV result is reproduced, but also the dependence of the spectrum on the matching scales disappears order by order in $1/k^2_\perp$. Again, these cancellations resemble the situation in the energy spectrum calculation~\cite{IOE3}. Additionally, the explicit and detailed calculation carried out in order to check that the IOE recovers the GLV result provides a non-trivial cross-check on the main formulas derived in this paper.

Regarding the final numerical evaluation we find that, for the plasma brick model with LHC-inspired parameters, the computing time is in the ballpark of the LO/BDMPS-Z result~\cite{ASW2}. More concretely, we have evaluated the code's performance and encountered that the small $\omega$ and small $\k$ regime requires more computational power due to the oscillating phases in the integrands. We provide ancillary {\tt Python} files with the IOE spectrum together with the GLV expressions in Ref.~\cite{python_git}. The comparison with an all-orders resumed spectrum reveals a globally good agreement between the NLO spectrum from the Improved Opacity Expansion and the full solution, for this set of medium parameters. In particular, the agreement improves at high-frequencies, where the deviations between the two approaches are below $10\%$. This is a remarkable result given the relative simplicity of the approach presented in this paper as compared to larger computational cost needed to resum the spectrum to all orders. It is indicative of the power of the IOE approach to capture the correct dynamics at small and large frequencies simultaneously. A more thorough comparison with the full numerical result including a scan of the parameters space is left for future work. We expect the agreement to systematically improve for denser or larger plasma for which the scale separation between $Q_r$ or  $Q_b$ and the infrared scale $\mu^2$ is larger. Obviously, the IOE scheme is exact asymptotically.

Our results provide for the first time a unified analytic framework for the fully differential medium-induced radiation spectrum that accounts for both the GLV and BDMPS limits. We expect that adopting our radiative kernel in future phenomenological studies would substantially reduce model dependence of jet quenching observables as well as theoretical uncertainties on the extraction of medium transport properties such as the jet quenching parameter, $\hat q$ that is a function of the typical scale of the process. Two phenomenological applications have been already proposed in the literature that concern quenching effects on the jet spectrum~\cite{Mehtar-Tani:2021fud,Takacs:2021bpv}. A natural continuation of this work is to use the in-medium radiative kernel that we have derived in this paper to analytically compute observables where the gluon transverse momentum information is not integrated out, namely jet substructure observables. In particular, on-going measurements of the $k_\perp$-distribution of the hardest splitting~\cite{Ehlers:2020piz} would benefit from a theoretical calculation in which both multiple soft scatterings and hard momentum exchanges are correctly incorporated. This study would open up a new theoretical window onto extending the IOE framework to describe the energy loss of a quark-antiquark antenna. In parallel, we would like to implement the formulas that we have derived in this paper in a suitable Monte Carlo framework such as Ref.~\cite{Saclay,Blanco:2020uzy}. 
 
\section*{Note}
While this manuscript was being produced an independent derivation of the differential spectrum (for a massive quark) using the IOE approach was presented in Ref.~\cite{Blok:2020jgo}. Although performing a numerical comparison is beyond the scope of this work we would like to point out a couple of differences. Firstly, in comparison with \cite{Blok:2020jgo} we have presented results for a generic medium profile for which we were able to reduce further the number of integration variables. Secondly,  in Ref.~\cite{Blok:2020jgo} a single matching scale was used for the radiative and broadening parts, i.e., $Q_b=Q_r$ which we found leads to an incorrect description of the spectrum.

\section*{Acknowledgements} 
We are grateful to the authors of Ref.~\cite{CarlotaFabioLiliana} for providing the numerical results from their study. In particular, we wish to thank Carlota Andrés and Fabio Dominguez for clarifying some important details in their work and for the careful reading of the present manuscript. We wish to thank Carlos Salgado for helpful discussions on related problems and Liliana Apolinário for providing clarifications regarding the GLV result obtained in Ref.~\cite{CarlotaFabioLiliana}.
The project that gave rise to these results received the support of a fellowship from ``la Caixa" Foundation (ID 100010434). The fellowship code is LCF/BQ/ DI18/11660057. This project has received funding from the European Union's Horizon 2020 research and innovation program under the Marie Sklodowska-Curie grant agreement No. 713673. J.B. is supported by Ministerio de Ciencia e Innovacion of Spain under project FPA2017-83814-P; Unidad de Excelencia Maria de Maetzu under project MDM-2016-0692; European research Council project ERC-2018-ADG-835105 YoctoLHC; and Xunta de Galicia (Conselleria de Educacion) and FEDER. The work of Y. M.-T. was supported by the U.S. Department of Energy, Office of Science, Office of Nuclear Physics, under contract No. DE- SC0012704.  K. T. is supported by a Starting Grant from Trond Mohn Foundation (BFS2018REK01) and the University of Bergen. Y. M.-T. acknowledges support from the RHIC Physics Fellow Program of the RIKEN BNL Research Center. A.S.O.’s work was supported by the European Research Council (ERC) under the European Union’s Horizon 2020 research and innovation programme (grant agreement No. 788223, PanScales).


\appendix 

\section{The analytic solutions of the emission kernel $\cK$ }\label{app:cK_appendix}

In this appendix, we discuss two analytic solutions for the emission kernel $\cK$ satisfying
\begin{equation}\label{eq:cK_Sch_app}
  \left[i\partial_{t_2}+\frac{\bdel^2_\x}{2\omega^2}+iv(\x,t_2)\right]\cK(\x,t_2;\y,t_1)=i\delta^{(2)}(\x-\y)\delta(t_2-t_1) \, .   
  \end{equation}
This propagator obeys a Dyson-like relation reading~\cite{book_Kleinert:2004ev}
\begin{equation}\label{eq:Dyson_GLV_app}
  \cK(\x,t_2;\y,t_1) = \cK_0(\x,t_2;\y,t_1)-\int_\z \int_{t_1}^{t_2}  \rmd s \, \cK_0(\x,t_2;\z,s)v(\z,s)\cK(\z,s;\y,t_1) \, ,
  \end{equation}
where $\cK_0$ corresponds to vacuum solution to Eq.~\eqref{eq:cK_Sch_app} with $v=0$. Alternatively, and as discussed in the main text, one can write an equivalent relation to \eqn{eq:Dyson_GLV_app}, but expanding around the solution to Eq.~\eqref{eq:cK_Sch_app} with $v\to \vLO$ and using the decomposition $v=\vLO+\delta v$ (see Eq.~\eqref{eq:v_IOE}),
\begin{equation}\label{eq:cK_ful_IOE_app}
  \cK(\x,t_2;\y,t_1) = \cK^{\rm LO}(\x,t_2;\y,t_1)-\int_\z \int_{t_1}^{t_2}  \rmd s \, \cK^{\rm LO}(\x,t_2;\z,s)\delta v(\z,s)\cK(\z,s;\y,t_1) \, .
  \end{equation}
Eqs.~\eqref{eq:Dyson_GLV_app} and \eqref{eq:cK_ful_IOE_app} are particularly useful since $\cK_0$ and $\cK^{\rm LO}$ admit a closed form, which can be obtained by directly solving Eq.~\eqref{eq:cK_Sch_app}, and thus they can be easily applied in a perturbative framework.

In the first case where $v=0$, $\cK_0$ is the Green's function of a Schr\"odinger equation describing the motion of a non-relativistic free particle in two dimensions, and thus reads
\begin{align}\label{eq:cK_vac_app}
\cK_0(\x,t_2;\y,t_1)=   \frac{\omega }{2\pi i (t_2-t_1) } \exp\left(i\frac{\omega(\x-\y)^2}{2(t_2-t_1)}\right) \, .
\end{align}
 
The case where $v(\x)=\vLO(\x)$ in Eq.~\eqref{eq:cK_Sch_app} is also easily solved since in this case $\cK$ is the Green's function associated to the motion of a single particle in a harmonic potential. For quadratic potentials, the exact solution to Eq.~\eqref{eq:cK_Sch_app} can be exactly obtained by using the so called method of fluctuations~\cite{book_Kleinert:2004ev,ASW2,IOE1,Arnold_simple}, resulting in
\begin{align}\label{eq:cK_BDMPS_app}
 \cK^{\rm LO}(\x,t_2;\y,t_1)&= \frac{\omega}{2\pi i S(t_2,t_1)}\exp\left( \frac{i\omega}{2S(t_2,t_1)} \left[ C(t_1,t_2)\,\x^2+C(t_2,t_1)\,\y^2-2 \x\cdot\y\right] \right) \,,
\end{align}
where we recall that the $C$ and $S$ functions satisfy
\begin{equation}\label{eq:Abel}
\begin{split}
&\left[\frac{\rmd^2}{\rmd^2t}+\Omega^2(t)\right]S(t,t_0)=0 \, ,\quad S(t_0,t_0)=0 \,,\quad \partial_t  S(t,t_0)_{t=t_0}=1 \, , \\   
&\left[\frac{\rmd^2}{\rmd^2t}+\Omega^2(t)\right]C(t,t_0)=0 \, ,\quad C(t_0,t_0)=1 \,,\quad \partial_t C(t,t_0)_{t=t_0}=0\, .
\end{split}
\end{equation}
For any $\Omega$, i.e. for a generic time profile of the medium, one can derive certain identities relating the $C$ and $S$ functions that were employed in the main text to simplify the emission spectrum. Firstly, these solutions are related by $C(t_1,t_2)=\partial_{t_2}S(t_2,t_1)$ and by the associated Wronskian ($W$), which reads
\begin{equation}
W=C(t_1,t_2)\partial_{t_1}S(t_1,t_2)-\partial_{t_1}C(t_1,t_2)S(t_1,t_2)=1\, ,    
\end{equation}
where we used the above the initial conditions. The condition $W=1$ can be used to show that 
\begin{equation}\label{eq:prop_app_1}
\partial_{t_1}\frac{C(t_1,t_2)}{S(t_1,t_2)}=-\frac{C(t_1,t_2)\partial_{t_1}S(t_1,t_2)-\partial_{t_1}C(t_1,t_2)S(t_1,t_2)}{S^2(t_1,t_2)} =-\frac{1}{S^2(t_1,t_2)} \, ,
\end{equation}
which is used to derive \eqn{eq:id_1}. In addition, $W=1$ implies that $C$ and $S$ are linearly independent solutions and thus any other solution to the above ordinary differential equation can be written as a linear combination of them. Using this fact, and for a time ordering $t_2>t_1>t_0$, any solution in $(t_2,t_1)$ can be written as~\cite{Arnold_simple}
\begin{equation}\label{eq:linear_rel_C_S}
\begin{split}
&S(t_2,t_1)=C(t_1,t_0)S(t_2,t_0)-S(t_1,t_0)C(t_2,t_0) \, ,\\ 
&C(t_2,t_1)=-\partial_{t_1}C(t_1,t_0)S(t_2,t_0)+\partial_{t_1}S(t_1,t_0)C(t_2,t_0) \, ,
\end{split}    
\end{equation}
where it is easy to verify that these equations satisfy the above initial conditions and that $S(t_2,t_1)=-S(t_1,t_2)$. This decomposition of the $C$ and $S$ is extensively used in the main text; here we give a simple application to derive another useful identity. 

Let us consider the brick model introduced in the main text, with a medium of extension $L$ such that  $\hat{q}(t\geq L)=0$, but still allowing the jet quenching parameter not to be constant in time inside the medium. In the vacuum, the solutions to the $C$ and $S$ functions are trivially found
\begin{equation}
\begin{split}
 S(t_2,t_1)=t_2-t_1 \, , \quad   C(t_2,t_1)=1 \, ,
\end{split}    
\end{equation}
and indeed when introduced back in Eq.~\eqref{eq:cK_BDMPS_app} give Eq.~\eqref{eq:cK_vac_app}. Using Eq.~\eqref{eq:linear_rel_C_S} with $t_2=+\infty$, $t_1>L$ and $t_0=L$, combined with the explicit forms for vacuum $C$ and $S$ solutions, one observes that terms proportional to $S(t_2,t_1)$ dominate, leading to the handy formula 
\begin{equation}
\frac{C(\infty,t_1)}{S(\infty,t_1)} = -\frac{\partial_{t_1} C(t_1,L)}{C(t_1,L)}=\Omega^2(t_1) \frac{S(t_1,L)}{C(t_1,L)} \, ,  
\end{equation}
where the last equality holds if $C$ is even in its arguments.
Finally, in the case of the brick model with a time independent $\hat{q}$ inside the medium, one finds that 
\begin{equation}
  \begin{split}
   S(t_2,t_1)=\frac{\sin(\Omega(t_2-t_1))}{\Omega} \, , \quad   C(t_2,t_1)=\cos(\Omega(t_2-t_1))\, .
  \end{split}    
  \end{equation}
Note that for the special cases of vacuum and static medium profiles, the $C$ function is even in its arguments.

\section{Solutions to the IOE matching scale}\label{app:Q}

In this appendix we pin down some of the basic properties of the real solutions to Eq.~\eqref{eq:Q_r}.~\footnote{As noted in Ref.~\cite{Takacs:2021bpv}, the transcendental equation can be written in terms of special functions, i.e. $Q^2_r=\mu^2_\ast\exp[-W_{-1}(2\mu^4_\ast/(\omega\hat q_0))/2]$, with $W_{i}(x)$ the Lambert function on the $i$-th branch.}
It is convenient to rewrite Eq.~\eqref{eq:Q_r} as a fixed point equation. Introducing

\begin{equation}
\label{eq:fpoint}
f_{\rm point}(x)\equiv x- \sqrt{\alpha \log x } \, ,
\end{equation}
where $x=Q^2/\mu_\ast^2 >1 $, $\alpha=\hat{q}_0\omega/\mu_\ast^4>0$.  

\begin{figure}[h!]
    \centering
    \includegraphics[scale=0.8]{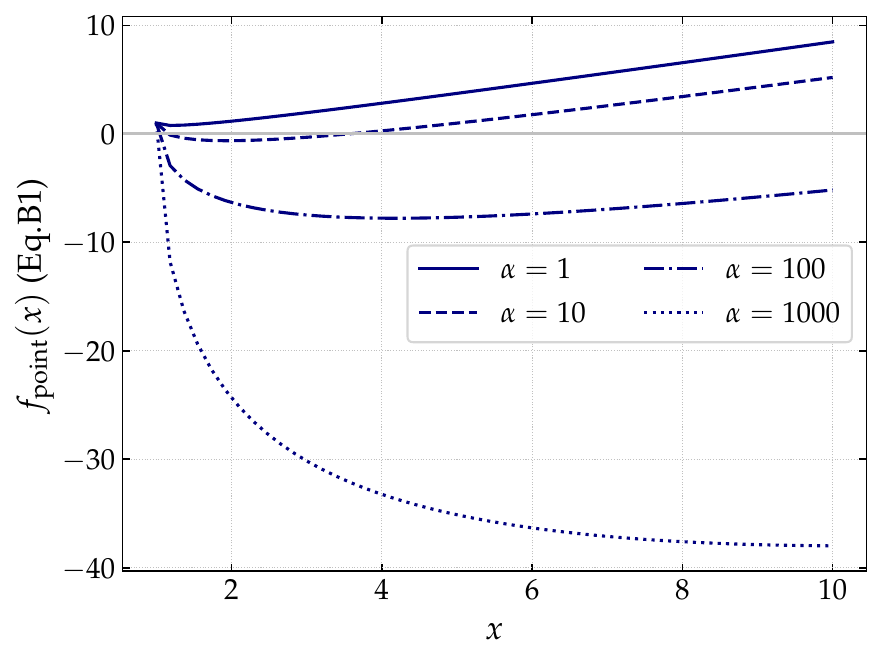}
    \caption{Function $f_{\rm point}$ given by Eq.~\eqref{eq:fpoint}  for several values of $\alpha$.}
    \label{fig:f_point}
\end{figure}

The function $f_{\rm point}$ is defined in the domain $[1,\infty)$ (avoiding the Bethe-Heitler region) and the values of $x$ for which $f_{\rm point}(x)=0$ correspond to the solutions to Eq.~\eqref{eq:Q_r}, after rescaling by $Q^2$. We notice that as $x\to \infty$, $f_{\rm point}\to \infty$ and $f_{\rm point}(1)=1$. Taking the first derivative of $f_{\rm point}$ and requiring the function to have its extremes at $y$, we find  
\begin{equation}
\log y=\frac{1}{4 \alpha y^2} \, ,
\end{equation}
which implies that the roots are in general not unique: either there is no real solution, there is a single solution or at most two real solutions. For a real solution to this equation to exist one must require that
\beq
\alpha > 2 \rme \iff \hat{q}_0\omega > 2 \rme \mu_\ast^4\, , 
\eeq
which can be derived from the limiting case where there is a single solution~\cite{Andres:2020kfg}. In particular we find that
\begin{equation}
f_{\rm point}(y)=\sqrt{\frac{\alpha}{4\log\left(y\right)}}-\frac{2\alpha}{y}  \, ,
\end{equation}
so that in the regime of interest where $\alpha \gg 1$, one always has 2 real solutions ($Q^-$ and $Q^+$) to \eqn{eq:Q_r}, which are numerically obtained depending on the initial condition used in solving the recursion relation. Since one typically uses as an initial condition $Q_0 \gg \mu_\ast$, the solution obtained is always the largest root $Q^+$, while as $\alpha \to \infty$, $Q^-\to 1$. These observations are illustrated in Fig.~\ref{fig:f_point}, where we numerically confirmed that in the limit $\alpha \gg 1$, there are always two roots (i.e. two real solutions for $Q$) with the smaller one asymptotically approaching $1$. In the opposite case, when $\alpha\ll 1$ there are no real solutions, as expected in the unphysical region where $\mu_\ast^2\gg \sqrt{\hat{q}_0\omega}$.

\section{Details on the adiabatic prescription and the vacuum contribution} \label{app:vacuum-derivation}
In this appendix, we discuss the vacuum limit of the full spectrum and the importance of the adiabatic prescription, ensuring that interactions are properly turned off at asymptotically large times. We also give a justification for the slightly modified adiabatic prescription that allows us to carry out time integrations for a general medium profile while correctly accounting for the vacuum contribution. 

\subsection{Revisiting the pure vacuum spectrum }\label{app:vacuum-derivation_details}
Before turning to the in-medium calculation, let us revisit the calculation of gluon emission in the vacuum, with the goal of understanding how to tame the emission of vacuum-like radiation at asymptotically large times. The matrix element which encodes the production of soft radiation with energy $\omega$ and transverse momentum $\k$ off a hard parton with light-cone energy $p^+$, reads
\beq\label{eq:vvv_1}
\cM \sim -ig (2p^+) \frac{\k }{(p+k)^2+i\epsilon } \simeq -ig (2p^+) \frac{\k }{2 p^+k^- +i\epsilon } \simeq -ig 2 \omega \frac{\k^i }{\k^2 } \,,
\eeq
where we have used $k^-=k_\perp^2/2\omega$ and anticipated that only the transverse component of the gluon will contribute at leading order. Squaring the amplitude and inserting the necessary phase space, spin and color factors, one recovers the vacuum spectrum introduced in \eqn{eq:spectrum-vac}
\beq\label{eq:vvv_2}
(2\pi)^2\omega\frac{\rmd I^{\rm vac}}{ \rmd \omega  \rmd^2 \k} =\frac{|\cM|^2C_R}{4 \pi \omega^2} = \frac{4\bar{\alpha}\pi}{k_\perp^2} \, . 
\eeq
To make contact with the main text calculation, we use the typical trick to write the denominator in an integral representation
\beq
 \frac{- i}{k^-+i\epsilon} = \int_0^\infty  \rmd t \,  \, \rme^{i \frac{k_\perp^2}{2\omega} t  - \epsilon t}\,.
\eeq
Going back to \eqn{eq:vvv_1}, we can express the square amplitude in terms of a double integral
\beq
 |\cM|^2   
 &=& g^2 k_\perp^2 2 \Re \int_0^\infty  \rmd t_2 \, \int_0^{t_2}\rmd t_1 \, \rme^{-i \frac{k_\perp^2}{2\omega} (t_2-t_1)  - \epsilon (t_2+t_1)}\, .
\eeq
Although this representation seems unnecessarily more complicated than the one in \eqn{eq:vvv_2}, it is of the form of the medium-induced spectrum in \eqn{eq:spectrum}. In addition, we observe that the adiabatic prescription we introduced in the main text is related to the Feynman prescription for the pole structure of the propagators.
The integral over $t_1$ is easy to carry out and gives
\beq\label{eq:vvv_3}
 |\cM|^2   
&=& g^2 k_\perp^2 2 \Re \int_0^\infty  \rmd t_2 \, \,\frac{i}{ \left(\frac{k_\perp^2}{2\omega}+i\epsilon\right)} \left[\rme^{-i \frac{k_\perp^2}{2\omega} t_2 - \epsilon t_2}- \rme^{- 2 \epsilon t_2}\right]\, .
\eeq
At this point, one could just perform the remaining $t_2$ integral and recover \eqn{eq:vvv_2}. However, since we know the correct result and this example is sufficiently simple, let us swap the implicit limit $\epsilon\to 0$ with the integral in \eqn{eq:vvv_2} first. Doing this we would obtain for the vacuum spectrum
\beq
(2\pi)^2\omega\frac{\rmd I^{\rm vac}}{ \rmd \omega  \rmd^2 \k} = \frac{8\bar{\alpha}\pi}{k_\perp^2} \, . 
\eeq
which is twice the result in  \eqn{eq:vvv_2}, thus showing that the limit does not commute with the integral. As a consequence, one must always perform first the integrations and only take the limit $\epsilon\to 0$ at the final step. Performing first the $t_2$ integral in  \eqn{eq:vvv_3}, we obtain
\beq\label{eq:vvv_4}
 |\cM|^2   
&=& g^2k_\perp^2 2 \Re  \left[ \,\frac{1}{ \left(\frac{k_\perp^2}{2\omega}+i\epsilon\right)^2} - \,\frac{i}{ 2\epsilon \left(\frac{k_\perp^2}{2\omega}+i\epsilon\right)^2}\right]\,. 
\eeq
Taking the limit $ \epsilon \to 0$ in the second term we obtain 
\beq 
\,-\frac{i}{2 \epsilon (\frac{k_\perp^2}{2\omega}+i\epsilon)^2}=- \frac{i}{2\epsilon}-\frac{2\omega^2} {k_\perp^4}\,+\cO(\epsilon) \, .
\eeq
The divergent term, $\sim \epsilon^{-1}$, is purely imaginary and does not contribute to the squared amplitude. Keeping only the real term and combining with the first term in \eqn{eq:vvv_4}, we obtain
\beq
 |\cM|^2   
&=&  \frac{8 g^2\omega^2}{k_\perp^2} - \frac{4g^2 \omega^2}{k_\perp^2} = \frac{4g^2\omega^2}{k_\perp^2} \, ,
\eeq
matching \eqn{eq:vvv_2} after including the necessary overall factors, as expected.

In the next section, we will detail how to deal with the adiabatic $\epsilon$ prescription in the case of in-medium emission. The strategy followed will be the same as the one detailed here, i.e. only take the $\epsilon\to 0$ limit at the end. However, the time integrals are not as simple to perform as in the vacuum case, so slightly more elaborate techniques are necessary, which we follow to describe.

\subsection{Details on going from \eqn{eq:spectrum} to \eqn{eq:spectrum-2}}\label{app:medium-derivation_details}

The full spectrum (including the vacuum contributions) reads
\begin{align}\label{eq:help_here}
(2\pi)^2 \omega \frac{\rmd I^{\rm \text{full}}}{\rmd \omega \, \rmd^2 \k} &= \lim_{\epsilon \to 0}\frac{2\bar \alpha \pi}{\omega^2} \rmR \int_0^\infty \rmd t_2 \int_0^{t_2} \rmd t_1 \int_{\x} \, \rme^{-\epsilon(t_2+t_1) } \rme^{-i \k\cdot \x} \cP(\x,\infty;t_2) \nn
&\times \bdel_\x \cdot \bdel_\y \cK(\x,t_2; \y ,t_1)_{\y=0} \,,
\end{align}
where we have introduced the adiabatic regulator $\epsilon$, and the limit $\epsilon \to 0$ has to be taken \emph{after} the integrations have been performed \cite{Wiedemann:1999fq,Wiedemann}. We have shown in the previous section that the $t_1$ and $t_2$ time integrals can be performed for pure vacuum radiation albeit with a careful treatment of interactions at infinity. In the medium, it is in general not possible to perform analytically both time integrals. However, we may use the fact that the primitive of $\bdel_\y  \cK(\x,t_2; \y ,t_1)$ with respect to $t_1$ is known and carry out the $t_1 $ integral, see \eqn{eq:id_1}. Note that the $\rme^{-\epsilon t_1}$ suppression factor stands in the way of the immediate application of \eqn{eq:id_1}. However, this is can be dealt with in two ways which we shall present in what follows. 

Let us first briefly outline the discussion presented in Ref.~\cite{Caucal:2020zcz}. Defining the function 
\beq
f(t_2,t_1) = \frac{1}{k_\perp^2} \int_{\x} \, \rme^{-i \k\cdot \x} \cP(\x,\infty;t_2)\bdel_\x \cdot \bdel_\y \cK(\x,t_2; \y ,t_1)_{\y=0} \,,
\eeq
it is straightforward to show that at late times $t_2>t_1 > T$, where $\cP(\x,\infty;t_2) = 1$ and $\cK(\x,\y) \approx \cK_0(\x,\y)$, the above function reduces to
\beq
f(t_2,t_1) \big|_{t_2>t_1 > T} \simeq \rme^{-i \frac{k_\perp^2}{2\omega} (t_2-t_1)} \, ,
\eeq
with $T$ an arbitrary time chosen to be much larger than the typical medium extension, i.e. $T\gg L$. Using this fact that at late times $f$ can be written as pure phase $\rme^{i \phi (t_2-t_1)}$, one can show that~\cite{Caucal:2020zcz}
\begin{align}
\lim_{\epsilon  \to 0} \rmR \int_0^\infty \rmd t_2 \int_0^{t_2} \rmd t_1 \, \rme^{-\epsilon(t_2+t_1)} f(t_2,t_1) &= \lim_{\epsilon  \to 0} \rmR \int_0^\infty \rmd t_2\,\rme^{-\epsilon t_2} \int_0^{t_2} \rmd t_1 \,  f(t_2,t_1)  - \frac{1}{2 \phi^2} \,,
\end{align}
where $\phi=-\frac{k_\perp^2}{2\omega}$.
Applying this to the full emission spectrum above, the subtraction term becomes the vacuum spectrum
\beq
(2\pi)^2 \omega \frac{\rmd I^{\rm vac}}{\rmd \omega \, \rmd^2\k} = \frac{4 \bar \alpha \pi}{k_\perp^2} \, ,
\eeq
leading to the final spectrum for the medium-induced spectrum considered in Eq.~\eqref{eq:spectrum-2}. 

An alternative strategy consists in absorbing the $\rme^{-\epsilon t_1}$ prescription in $\cK$ while preserving its integrability by following the chain of operations:
\begin{align}
\lim_{\epsilon \to 0}\int_0^{\infty}\rmd t_2  \int_0^{t_2}\rmd t_1   &\cK_0(t_2-t_1|\omega) \rme^{-\epsilon (t_2+t_1)} \equiv   \lim_{\epsilon \to 0} \int_0^{\infty}\rmd t_2  \int_0^{t_2}\rmd t_1   \rme^{-i \frac{k_\perp^2}{2\omega}(t_2-t_1)} \rme^{-\epsilon (t_2+t_1)} \nn
&= \lim_{\epsilon' \to 0} \int_0^{\infty}\rmd t_2  \int_0^{t_2}\rmd t_1   \rme^{-i \frac{k_\perp^2}{2(\omega-i \epsilon')}(t_2-t_1)} \rme^{-2 \epsilon' \frac{k_\perp^2 }{2\omega } t_2} \nn
&=\lim_{\epsilon' \to 0} \int_0^{\infty}\rmd t_2  \int_0^{t_2}\rmd t_1   \cK_0(t_2-t_1|\omega-i\epsilon') \rme^{-2 \epsilon' \frac{k_\perp^2 }{2\omega } t_2} \, ,
\end{align}
where $\epsilon' =  \epsilon (2\omega/k_\perp^2)$. This shows that instead of using the $\rme^{-\epsilon t_1}$ adiabatic phase, one can do a slight rotation of the frequency $\omega$ in the complex plane. The generalization of the above discussion to the full $\cK$ is straightforward since the $\epsilon$ prescription is only relevant at asymptotically large $t_2$ and $t_1$. Let us show how this prescription applies to Eq.~\eqref{eq:spectrum}. The integral over $t_1$ reads
\begin{align}\label{eq:t1-int}
\int_0^{t_2} \rmd t_1 \, &\rme^{-\epsilon t_1} \partial_\y\,  \cK^{\rm LO}(\x,t_2;\y=0,t_1|\omega ) \, \to \,  \int_0^{t_2} \rmd t_1 \, \partial_\y \cK^{\rm LO}(\x,t_2;\y=0,t_1|\omega'=\omega-i \epsilon' )\nn
   &=\lim_{t_0\to 0}\frac{\omega'}{\pi i} \frac{\x}{\x^2} \, \left[ \rme^{\frac{i\omega' }{2} {\rm Cot}( t_2,0) \x^2} -\rme^{\frac{i\omega' }{2} {\rm Cot}( t_2,t_2-t_0) \x^2}\right]  \, \nn
   &=\frac{\omega'}{\pi i} \frac{\x}{\x^2} \, \rme^{\frac{i\omega' }{2} {\rm Cot}( t_2,0) \x^2}   \, ,
  \end{align} 
where we have used Eq.~\eqref{eq:prop_app_1} (see also \eqn{eq:id_1}) and in the last line we have used that for vanishing $t_0$,  ${\rm Cot}( t_2,t_2-t_0)  \to  1/t_0$, and the second term can thus be neglected. Using \eqn{eq:t1-int}, we obtain that the respective contribution to the spectrum reads
\begin{align}\label{eq:lol_2}
  & \lim_{\epsilon \to 0}\frac{-2 \bar{\alpha}}{\omega^2} \Re  \bigg[i\omega' \, \int_0^\infty \rmd t_2\, \rme^{-2\epsilon t_2} \int_\x \rme^{-i \k \cdot \x}\, \cP(\x,\infty;t_2)  
    \bdel_\x \left( \frac{\x}{\x^2} \, \rme^{\frac{i\omega' }{2} {\rm Cot}( t_2,0) \x^2}\right) \bigg] \, .
 \end{align}
When the differential operator acts on the exponential factor one recovers the medium induced contribution given for example in \eqn{eq:HO_any_med_1a}; in this appendix we are interested in the first term, which was overlooked before. It is easy to show that 
\begin{align}\label{eq:final_11}
\partial_\x \left(\frac{\x}{\x^2}\right) = 2\pi \,  \delta^{(2)}(\x) \, ,
\end{align}
is a valid representation for the Dirac delta function. Using this result, the $\x$ integral in \eqn{eq:lol_2} becomes trivial. After some simple manipulations, we obtain that the net contribution to the spectrum is simply
 \begin{align}\label{eq:lol_3}
  \lim_{\epsilon \to 0}- \frac{4 \pi \bar{\alpha}}{\omega^2}  \Re  \bigg[\,i \omega' \,  \int_0^\infty \rmd t_2\, \rme^{-2 \epsilon  t_2}  \bigg] =  - \frac{2\pi \bar{\alpha}}{\omega^2 \epsilon }  \Re  \bigg[\,i \left(\omega -  i\epsilon \frac{2\omega }{k_\perp^2}\right)\bigg] =  - \frac{4\pi \bar{\alpha}}{k_\perp^2}    \, .
    \end{align}
    This contribution exactly matches the term subtracted in \eqn{eq:spectrum-2}. Also notice that as a consequence, in the main text calculation, one can ignore the action of any differential operator on $\frac{\x}{\x^2}$, since its contribution has already been taken into account.

Note that the prescription $\epsilon'$ is only relevant for the purely vacuum term emerging from \eqn{eq:lol_2}. In the term considered in the main text, one can safely set $\omega' \to \omega$ and $\epsilon' \to \epsilon$, owing to the fact that there is only one remaining time integral $t_2$ for which the details of the regularization, such as multiplying $\epsilon$ by an arbitrary factor, do not matter. In addition, one can show that the adiabatic prescription can be ignored for higher order contributions in the IOE. A simple way to see this, at NLO, is to introduce \eqn{eq:mmmm_1} and \eqn{eq:mmmm_2} in \eqn{eq:t1-int}, after adjusting the time integration limits and inserting extra $\cK^{\rm LO}$ factors inherited from the full NLO spectrum; see \eqn{eq:spectrum-ioe-nlo}. Also, one must recall that the NLO contributions are proportional to $\delta v$. The calculation follows as for the above example, where the terms coming from  $\partial_\x \left(\frac{\x}{\x^2}\right)$, will be proportional to $\delta v(\x) \delta^{(2)}(\x)= 0$, showing that the adiabatic regularization can be overlooked at higher orders in the IOE.


\section{The \textbf{in-out} contribution to the IOE spectrum in the multiple soft scattering regime}\label{app:In_Out_example}
In this appendix, we explicitly show that in the multiple soft regime considered in Section~\ref{sec:MS_IOE}, the NLO contribution in the IOE scales as the LO term multiplied by a logarithm that arises from the ratio $\frac{\hat{q}_0}{\hat{q}}$. For the sake of the argument, we focus on the \textbf{in-out} contribution, which although giving being sub-leading, still exhibits this scaling and it is straightforward to compute. In this kinematic regime $\Omega L\gg 1$, and we can use 
\begin{equation}\label{eq:help1}
\displaystyle\lim_{\frac{\omega}{\omega_c} \to 0} \Omega \cot (\Omega L)={\rm i}\Omega    \, ,
\end{equation}
together with $\k^2\ll \sqrt{\hat{q}\omega}\sim \Omega \omega $, to write Eq.~\eqref{eq:HO_brick_out_2} as
\begin{equation}\label{eq:LO_smallwk_IO}
(2\pi)^2\omega\frac{\rmd I^{\rm LO}_{\rm \text{in-out}}}{\rmd \omega \rmd^2 \k}=  -\frac{8\Bar{\alpha}\pi}{\sqrt{\hat{q}\omega }} + \mathcal{O}(k_\perp^{-2})\, .
\end{equation}
For the NLO part, we use the small argument expansion of ${\rm Ei}(x)\approx \gamma_E+\log x$ to reduce Eq.~\eqref{eq:bbS_beta} to

\begin{equation}\label{eq:smallk_S2}
I_b(x,y)\approx 4x^2 \log \frac{y}{4x E_\gamma} \, .
\end{equation}
Plugging this result into Eq.~\eqref{eq:In_out_NLO_final} yields
\begin{align}
&(2\pi)^2\omega\frac{\rmd I^{\rm NLO}_{\rm \text{in-out}}}{\rmd \omega  \rmd^2 \k}=8\Bar{\alpha}\hat{q}_0\pi \Re \bigg[\int_0^L \rmd t_1 \, \frac{C_{12}^2}{\jhat_{21}^2}\log \frac{\jhat_{21}}{4Q_r^2E_\gamma C_{12}^2}\bigg]
\nn
&= \frac{8\Bar{\alpha}\pi}{\sqrt{\hat{q}_0 \omega \log^3\left(\frac{Q_r^2}{\mu_{\star}^2}\right)}}
\Re \bigg[i\int_0^\Phi \rmd u \, 
\frac{\log\left((-1+i)\frac{4E_\gamma Q_r^2}{\sqrt{\hat{q}\omega}}\frac{1}{\cot((1-i)u)-\tan((1-i)(\Phi-u))}\right)}{(\cot((1-i)u)\cos((1-i)(\Phi-u))-\sin((1-i)(\Phi-u)))^2}
\bigg] \, ,
\end{align}
where $E_\gamma\equiv\exp(1-\gamma_E)$ and $\Phi=L\sqrt{\frac{\hat{q}}{4\omega}}\gg1$. Taking the upper limits in the integrals to infinity, using Eq.~\eqref{eq:help1} and splitting the logarithm in the integrand, one can use the following numerical integrals,\footnote{In the first integral one could further approximate the tangent inside the logarithm and obtain instead
\begin{equation*}
  i \int_0^\infty \rmd u \, \frac{\log\left((1+i)2E_\gamma\right)}{(i\cos((1-i)u)-\sin((1-i)u))^2}\approx-0.16928 - 0.56198  \, i \, ,
  \end{equation*}
but for the current argument these small numerical differences are not important.
   }
   \begin{equation}\label{eq:smal_wk_int1}
i \int_0^\infty \rmd u \, \frac{\log\left((-1+i)4E_\gamma\frac{1}{i-\tan((1-i)u)}\right)}{(i\cos((1-i)u)-\sin((1-i)u))^2}\approx-0.26585-0.65855 \, i \, ,
   \end{equation}
and
\begin{equation}\label{eq:smal_wk_int2}
i \int_0^\infty  \rmd u \, \frac{1}{(i\cos((1-i)u)-\sin((1-i)u))^2}=-\frac{1}{4}(1+i) \, ,
\end{equation}
to find that the spectrum is approximately given in this regime by
\begin{equation}\label{eq:IOE_smallkt_final}
(2\pi)^2\omega\frac{\rmd I^{\rm NLO}_{\rm \text{in-out}}}{\rmd \omega \rmd^2 \k}\approx \left((2\pi)^2\omega\frac{\rmd I^{\rm LO}_{\rm \text{in-out}}}{\rmd \omega \rmd^2 \k}\right) \, \frac{\hat{q}_0}{\hat{q}} \left(0.26(5)+\frac{1}{4}\log \frac{Q_r^2}{\sqrt{\hat{q}\omega}}\right) \, .
\end{equation}
The logarithm inside the brackets is small~\cite{IOE3} since at leading-logarithmic order $Q_r^2=\sqrt{\hat{q}\omega}$, as discussed in Section~\ref{sec:energy-spectrum}. As detailed above, indeed we explicitly find that the NLO result scales as the LO term times a logarithmic factor encapsulated by $\frac{\hat{q}_0}{\hat{q}}$.

\bibliographystyle{elsarticle-num}

\bibliography{Lib.bib}

\end{document}